 \documentclass[final,5p,times]{elsarticle}
\usepackage{amsfonts}
\usepackage{amsmath}
\usepackage{systeme}

\usepackage{mathabx}						% Cappeli e accenti
\usepackage{graphicx}
\usepackage{subcaption}
\usepackage{caption}
\usepackage{tikz,pgfplots}
\pgfplotsset{compat=1.17}
\usepackage{epsfig}
\usepackage{mathtools}
\usepackage{booktabs}                       % Tabelle
\usepackage{tabularx}                       % tabelle di larghezza prefissata
\graphicspath{{FIGURES/}}                   % cartella dove sono riposte le immagini
\usepackage{accents}                        % Cerchio sopra le lettere
\usepackage{dblfloatfix}                    % FBottom figure in a two-column document
\usepackage{supertabular}
\usepackage{multirow}
\usepackage{longtable}
\usepackage{pdflscape}
\usepackage{afterpage}
\usepackage{placeins}
\usepackage{capt-of}% or use the larger `caption` package
\usepackage{comment}                        % commenti
\usepackage{multicol}
    \makeatletter
\newsavebox{\bigleftbox}

\newcounter{subassumption}[asu]

\makeatletter
\renewcommand{\p@subassumption}{\theasu}% Counter prefix.
\makeatother

\makeatletter
\def\BState{\State\hskip-\ALG@thistlm}
\makeatother

\usepackage[colorlinks=true,urlcolor=blue]{hyperref} 

\journal{Applied Thermal Engineering}

\usepackage{algorithm}% <=================http://ctan.org/pkg/algorithms
\usepackage{algpseudocode}% <========== http://ctan.org/pkg/algorithmicx

\begin{document}

\begin{frontmatter}

%% Title, authors and addresses
%% use the tnoteref command within \title for footnotes;
%% use the tnotetext command for theassociated footnote;
%% use the fnref command within \author or \address for footnotes;
%% use the fntext command for theassociated footnote;
%% use the corref command within \author for corresponding author footnotes;
%% use the cortext command for theassociated footnote;
%% use the ead command for the email address,
%% and the form \ead[url] for the home page:
%% \title{Title\tnoteref{label1}}
%% \tnotetext[label1]{}
%% \author{Name\corref{cor1}\fnref{label2}}
%% \ead{email address}
%% \ead[url]{home page}
%% \fntext[label2]{}
%% \cortext[cor1]{}
%% \address{Address\fnref{label3}}
%% \fntext[label3]{}

\title{Experimental analysis of heat and mass transfer in non-isothermal sloshing using a model-based inverse method}
%% use optional labels to link authors explicitly to addresses:
%% \author[label1,label2]{}
%% \address[label1]{}
%% \address[label2]{}

\author[vki,ulb]{Pedro A. Marques\corref{cor1}}
% \author[uchi]{Laura Peveroni\corref{cor1}}
\author[vki]{Alessia Simonini}
\author[vki]{Laura Peveroni}
% \ead{apasculli@unich.it}
\author[vki]{Miguel A. Mendez}

\cortext[cor1]{Corresponding author.\\ E-mail address: \href{pedro.marques@vki.ac.be}{pedro.marques@vki.ac.be}}
%\cortext[cor2]{Principal corresponding author}
%\fntext[fn1]{This is the specimen author footnote.}
%\fntext[fn2]{Another author footnote, but a little more longer.}
%\fntext[fn3]{Yet another author footnote. Indeed, you can have
%any number of author footnotes.}

% \address[uchi]{University G. D'Annunzio, Dept. of Engineering Geology (INGEO), Chieti-Pescara, Italy}
\address[vki]{von Karman Institute for Fluid Dynamics, Waterloosesteenweg 72, 1640 Sint-Genesius-Rode, Belgium}
\address[ulb]{Université Libre de Bruxelles, Av. Franklin Roosevelt 50, 1050 Bruxelles, Belgium}

\begin{abstract}
%% Text of abstract

Nonisothermal liquid sloshing in partially filled reservoirs can significantly enhance heat and mass transfer between liquid and ullage gasses. This can result in large temperature and pressure fluctuations, producing thrust oscillations in spacecraft and challenging thermal management control systems. This work presents an experimental characterization of the thermodynamic evolution of a cylindrical reservoir undergoing sloshing-induced thermal de-stratification. We use a 0D model-based inverse method to retrieve the heat and mass transfer coefficients in planar and swirl sloshing conditions from the temperature and pressure measurements in the liquid and the ullage gas. The experiments were carried out in the SHAKESPEARE shaking table of the von Karman Institute in a cuboid quartz cell with a cylindrical cut-out of 80 mm diameter in the centre, filled up to 60mm with the cryogenic replacement fluid HFE-7200. A thermal stratification with $\Delta T=25$ K difference between the ullage gas and liquid was set as the initial conditions. 
A pressure drop of 90\% in the ullage gas was documented in swirling conditions. Despite its simplicity, the model could predict the system's thermodynamic evolution once the proper transfer coefficients were derived.

%---------------------------
\end{abstract}


\end{frontmatter}

%\section*{Highlights}

%Main outcomes of the paper in 3-5 bullet points (max 85 characters per bullet):

%\begin{itemize}
 %   \item Planar sloshing mixed the flow around the interface, not reaching the bulk liquid. % 82 char
 %   \item Swirl sloshing fully mixed the liquid's thermal field, yielding a 90\% pressure drop. % 85 char
 %   \item 0D model reproduced transient and quasi-steady temperature and pressure evolutions. % 85 char
 %   \item Link between transient heat transfer coefficients and transient sloshing motion. % 85 char
%\end{itemize}

%======================================================%
%%%%%%%%%%%%%%%%%%%%%%%%%%%%%%%%%%%%%%%%%%%%%%%%%%%%%%%%
%%%%%%%%%%%%%%%%%%%%% INTRODUCTION %%%%%%%%%%%%%%%%%%%%%
%%%%%%%%%%%%%%%%%%%%%%%%%%%%%%%%%%%%%%%%%%%%%%%%%%%%%%%%
%======================================================%
\section{Introduction}
\label{sec:intro}

Cryogenic fuels such as LH$_2$, LCH$_4$ and LNG are gaseous in ambient conditions but must often be transported and stored in the liquid state to have cost-competitive volumetric energy densities \cite{Petitpas2018}. This requires challenging cryogenic conditions, especially for hydrogen, which can only exist as a liquid (LH$_2$) under 33 K. While these fuels are already employed in aerospace propulsion \cite{Braeunig2008}, significant technological advances are still required to make them economically viable for other applications on a global scale \cite{Ball2009,Kim2019,Dawood2020}.

The efficient storage of cryogenic liquids requires complex thermal management systems. Since no insulating system can entirely prevent heat exchange with the surrounding \cite{Khemis2004}, it is impossible to prevent some of the fuel from boiling off and thus the tank pressure from rising over time \cite{Sousa2017, Zuo2021, Jeon2021}. The simplest way to control the tank pressure is venting, but this brings a high cost of energy and mass loss. Additional liquefaction steps can be used to recover the vented fuel, but this further increase the operational costs.

Thermal losses increase significantly in fuel tanks installed on vehicles, whether these travel on sea, land, air, or space because external accelerations induce sloshing. 
Besides challenging the vehicle's stability and manoeuvrability \cite{Kajon2014}, sloshing can significantly increase heat and mass transfer rate between liquid and ullage gasses, leading to significant variations of the tank's pressure \cite{Moran1994,Liu2019, Liu2020}.

In typical operating conditions, the ullage gasses tend to be warmer than the liquid. This is due to the pressurization, typically up to $2-12$ bar \cite{Petitpas2018,Joseph2016}, and the heat fluxes with the environment \cite{Kinefuchi2020,Perez2021}. As a result, thermal stratification is produced in the tank if this is left in quiescent conditions. Sloshing produces a thermal mixing that can level the stratification and suddenly reduce the temperature at the gas-liquid interface \cite{Lacapere2009}. Since the interface naturally sets the saturation condition of the tank, a decrease in its temperature produces a decrease in the saturation pressure, hence condensation \cite{Arndt2011}. This is known in the literature as the `pressure drop effect' \cite{Lacapere2009,Behruzi2006}.

The opposite mechanism can also occur when heat exchange across the wall increases the interface temperature near the contact line and thus enhances evaporation \cite{Dreyer2007}, increasing the tank's pressure 
\cite{Schmitt2016}. The pressure fluctuations due to sloshing-induced evaporation or condensation can produce thrust oscillation in a spacecraft and force thermal management systems to react with venting or re-pressurization; both operations bring a cost that must be minimized.

The literature on the sloshing phenomenon is abundant and has been significantly fostered by the aerospace industry during the 1960s \cite{Bauer1964,Abramson1966,Lance1966}. These works focused on the (isothermal) wave dynamics and the damping rates of several fluids under many excitation conditions and tank shapes. More recently, the interest in the pressure drop effect has shifted towards characterizing the non-isothermal consequences of sloshing \cite{Moran1994, Lacapere2009,Arndt2011,Schmitt2016,Himeno2011,Kulev2014,Behruzi2019}. These works have focused on the thermodynamic response of partially filled tanks under different sloshing regimes, pressurization techniques, pressurant gasses and tank designs (e.g. with or without baffles). The data retrieved in these experiments allowed several authors to develop and validate simplified models to predict the thermodynamic changes in fuel tanks due to sloshing. The simplest and yet most popular approach is a quasi-dimensional (0D) formulation \cite{Petitpas2018, Osipov2011, Migliore2016, Grotle2018} in which the fuel tank system is composed of three control volumes (CVs): the ullage gas, the liquid, and the solid walls. Under this lumped formulation, the model equations are derived from each volume's energy and mass balances, using mass-averaged thermodynamic properties for each volume. This results in a coupled system of Ordinary Differential Equations (ODEs) which can be solved for a given set of inputs and initial states to predict evaporation and condensation rate and hence the thermodynamic evolution of the fuel tank. Since these models are simple to derive and computationally inexpensive to evaluate, they can be easily integrated with other sub-models to perform system simulations \cite{Daigle2011} in real time. 

However, these models must rely on closure laws for heat and mass transfer coefficients to close the system of equations. While several empirical correlations have been proposed for stationary/long-term storage conditions \cite{Joseph2016,Osipov2011, Migliore2016}, their estimation during sloshing events remains an open problem. Moreover, no established methodology has been proposed in the literature for driving these closure laws in dynamic conditions.

A natural approach is to `tune' these coefficients such that numerical prediction match well with experimental data (see \cite{Petitpas2018} and \cite{Grotle2018}). This can provide useful orders of magnitude, but it can hardly account for the time variation of these coefficients. To the authors' knowledge, the only correlation currently available for modelling the sloshing-enhanced thermal mixing has been proposed by Ludwig \& Dreyer \cite{Ludwig2013}. These authors have proposed an empirical law for the sloshing-induced `effective diffusivity', which depends on the system's initial condition and the maximum pressure variation measured in the tank. Their correlation, expressed in terms of a sloshing-based Nusselt number, proved valid for their experimental data and the experiments of Das \& Hopfinger \cite{Das2009} and Moran \textit{et. al} \cite{Moran1994}. However, the agreement was limited to the initial stages of the sloshing excitation.

This work presents an experimental characterization of the sloshing-induced heat and mass transfer and proposes a model-based inverse method to derive empirical correlations from the measurements of pressure and temperatures in the tank. The inverse method sets an optimization problem in which the closure terms are automatically tuned to minimize the discrepancy between experimental data and a model's prediction. This data-driven approach, common in data-assimilation literature \cite{Suzuki2015}, has been used in the calibration of constitutive laws \cite{Mendez2021}, modelling of dynamic menisci \cite{Fiorini2022}, thermal turbulence closure \cite{Fiore2022}, flow control \cite{Pino2022} and modelling closure for stagnation line flow in hypersonic reentry \cite{Gkimisis2022}.

This work gives a promising proof of concept on using such a model-based inverse method to derive empirical laws for slosh\-ing-induced heat and mass transfer from pressure and temperature measurements in the tank. We consider a reduced-scale upright cylindrical tank using the non-cryogenic replacement fluid HFE-7200 \cite{HFE7200_3M}. 
The experiments were carried out under different lateral sloshing excitation in the presence of an axial thermal stratification. The underlying thermodynamic model is a classic 0D formulation of the energy and mass balance between three control volumes (liquid, ullage gas and solid), while the optimization was carried out with the Basinhopping algorithm, complemented with a Monte Carlo approach for estimating model uncertainties.

The rest of the article is structured as follows.  Section \ref{sec:problem_description} presents an overview of the physical problem alongside all modelling assumptions and approximations. Section \ref{sec:operating_conditions} summarizes the experimental test cases and the theoretical aspects used to reproduce different sloshing conditions in the laboratory. Section \ref{sec:methodology} reports on the methodology, including a detailed descriptions of the experimental setup and procedure, the 0D model, and the inverse method. The results are shown and discussed in Section \ref{sec:results}. Conclusions and future outlook are presented in Section \ref{sec:conclusion}.

% ----------------------------------------
% JUSTIFICATION FOR ONLY LATERAL SLOSHING:
% ----------------------------------------
% This work aims to characterize and to model non-isothermal sloshing for space propulsion applications. While cryogenic propellants can experience different types of excitations depending on the flight stage, only the `propulsive phase' was considered. This is characterized by a gravity-dominated state in which sloshing is predominantly triggered by disturbances applied in the lateral direction \cite{Dreyer2007,Arndt2011}.
% ----------------------------------------
% ----------------------------------------

%======================================================%
%%%%%%%%%%%%%%%%%%%%%%%%%%%%%%%%%%%%%%%%%%%%%%%%%%%%%%%%
%%%%%%%%%%%%%%%%%%%%% PROBLEM SET %%%%%%%%%%%%%%%%%%%%%%
%%%%%%%%%%%%%%%%%%%%%%%%%%%%%%%%%%%%%%%%%%%%%%%%%%%%%%%%
%======================================================%

\section{Problem Statement and Objectives}
\label{sec:problem_description}

\begin{figure}[h]
	\centering
	\includegraphics[width=0.99\linewidth]{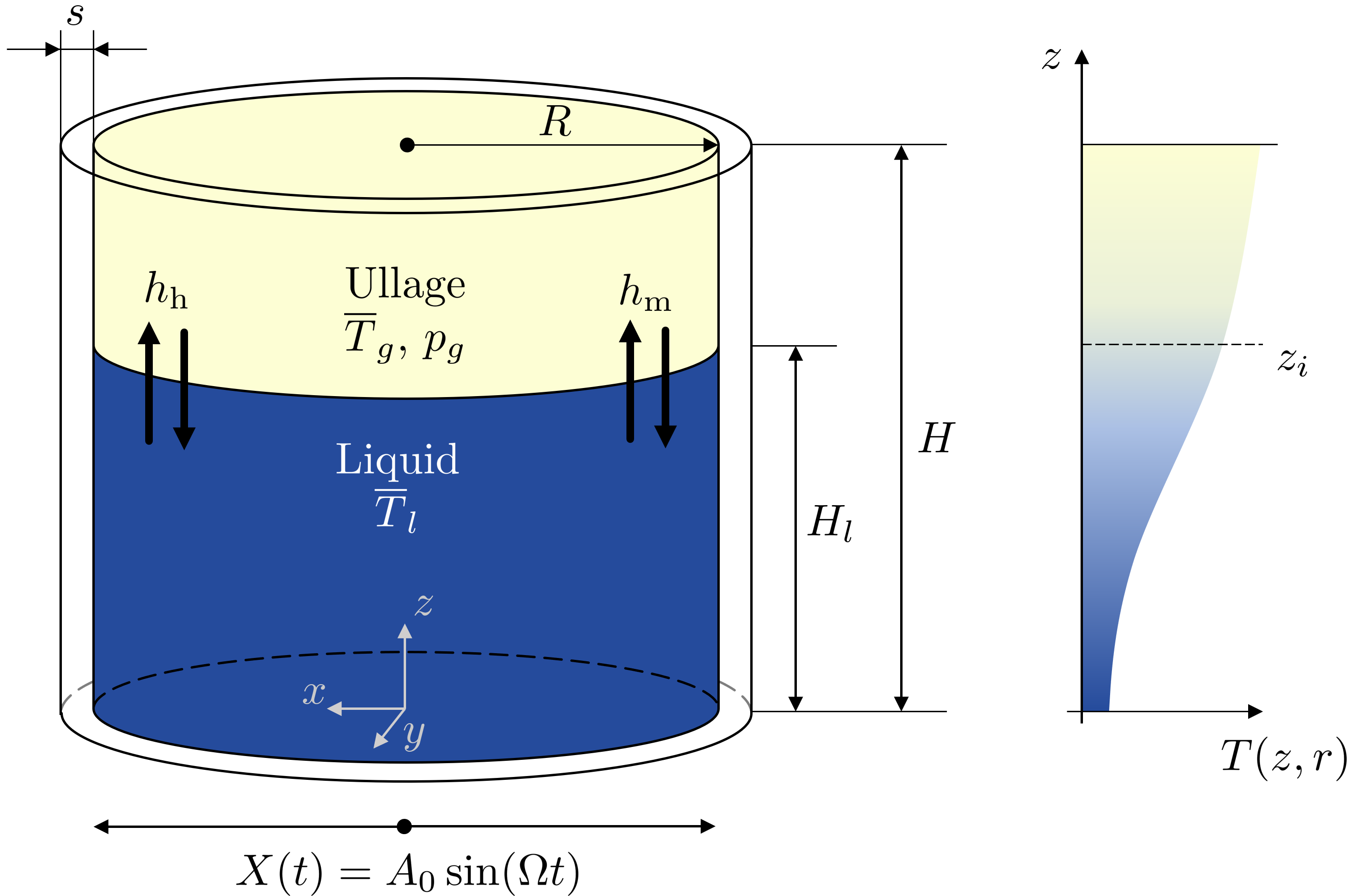}
	\caption{Schematic of the the problem set. Fuel tank with radius $R$, total height $H$, solid wall thickness $s$, and filled with liquid up to $H_l$ (left). Axial temperature distribution after pressurizing the tank at an arbitrary radial coordinate $r$ (right). Because the tank is pressurized up to a certain pressure $p_g$, the gas is initially warmer than the liquid.}
	\label{fig:problem_description}
\end{figure}

The problem considered in this work is schematically illustrated Figure \ref{fig:problem_description}. A cylindrical tank of radius $R$ and height $H$, with a flat top and bottom surfaces, is filled with liquid until a height $H_l$. The tank is closed and pressurized until the ullage gas is at a pressure $p_g$. The ullage is composed of a mixture of two species: an inert gas at partial pressure $p_a$ and the liquid's vapour at partial pressure $p_v$.

The tank walls are adiabatic with mass $m_w$ and specific heat capacity $c_w$. Because of the tank's pressurization, the system tends towards a thermally stratified configuration in which the liquid is colder than the gas. More specifically, to a first approximation, one might expect an axisymmetric temperature field $T(r,z)$, with $r$ the radial coordinate and $z$ the vertical one, as sketched in the right-hand side of Figure \ref{fig:problem_description}. Subscripts $l$, $g$ $w$ denote variables linked to the liquid, the gas and the walls. The gas-liquid interface is at the saturation temperature $T_\text{sat}(p_v)$ corresponding to the gas pressure, and a positive temperature gradient $\partial_z T$ develops at the interface (cf. Figure \ref{fig:problem_description}). No mass transfer occurs between liquid and gas if the ullage gas is saturated with the liquid vapour. 

In this work, we focus on the perturbation of this state due to lateral sloshing of the tank. We consider an harmonic displacement along $x$ with amplitude $A_0$ and angular frequency $\Omega$:

\begin{equation}
	X(t) = A_0 \sin(\Omega t).
\end{equation}

Sloshing induces the mixing of the liquid interface, thus a destratification of the thermal field. This reduces the (average) interface temperature $T_i$, which in turn leads to an increase in the heat transfer and mass transfer at the interface. Both mechanisms lead to a pressure drop. We here seek to analyze experimentally whether a simple 0D model of the heat and mass transfer can describe the thermodynamic evolution of the ullage gas after a sloshing event. Furthermore, let 
\begin{equation}
	\small
	\label{eq:heat}
	h_{li}(t)=\frac{\dot{Q}_{li}(t)}{S (\overline{T}_l(t)-T_i(t))}
	\,\,, \,\, 
	h_{gi}(t)=\frac{\dot{Q}_{gi}(t)}{S (\overline{T}_g(t)-T_i(t))}
\end{equation}

\begin{equation}
	\small
	\label{eq:heat_tank}
	h_{wl}(t)=\frac{\dot{Q}_{wl}(t)}{S_{wl} (\overline{T}_w(t)-T_l(t))}
	\,\,, \,\, 
	h_{wg}(t)=\frac{\dot{Q}_{wg}(t)}{S_{wg} (\overline{T}_w(t)-T_g(t))}
\end{equation}

\noindent
denote the instantaneous global heat transfer coefficients between the fluid phases and interface, and between the fluids and the tank walls, respectively, with $\dot{Q}$ the heat exchanged between the different regions, identified by their $l$ (liquid), $g$ (gas), $i$ (interface), $w$ (walls) subscripts. 
In these expressions, $S$ is the tank's cross section area, $S_{wl}$ and $S_{wg}$ the tank's wetted areas in contact with the liquid and gas at rest, $\overline{T}_l(t)$, $\overline{T}_g(t)$, $\overline{T}_w(t)$ the (instantaneous) volume averaged temperature in the liquid, gas, and solid walls, $T_i(t)$ the instantaneous spatial average of the interface temperature.
Let 

\begin{equation}
	\label{eq:mass}
	h_m(t)=\frac{\dot{m}_i(t)}{S (\rho_{v,\text{sat}}(t)-\overline{\rho}_{v}(t))}
\end{equation} denote the instantaneous global mass transfer coefficient between liquid and gas, with $\dot{m}_i$ the exchanged vapour mass flow rate, $\rho_{v,sat}$ the vapour density at the saturation temperature and $\overline{\rho}_v$ the volume average of the vapour density.

We seek to analyze whether a model-based inverse method can be used to retrieve $h_{li}$, $h_{gi}$ and $h_m$ from real-time measurements of pressure and temperature in the tank.

%======================================================%
%%%%%%%%%%%%%%%%%%%%%%%%%%%%%%%%%%%%%%%%%%%%%%%%%%%%%%%%
%%%%%%%%%%%%%%%%% SLOSHING CONDITIONS %%%%%%%%%%%%%%%%%%
%%%%%%%%%%%%%%%%%%%%%%%%%%%%%%%%%%%%%%%%%%%%%%%%%%%%%%%%
%======================================================%

\section{Sloshing conditions}
\label{sec:operating_conditions}

This section describes the excitation conditions applied to the fuel tank to produce liquid sloshing. The system was excited in the vicinity of the natural frequency corresponding to the first asymmetrical lateral sloshing mode. The natural frequencies in a partly-filled upright cylindrical reservoir subject to these conditions are given by:

\begin{equation}
	\label{eq:natural_freq}
	\omega_{mn}^2 =
	\left(
	\frac{g \xi_{mn}}{R}
	+
	\frac{\sigma}{\rho_l}
	\frac{\xi_{mn}^3}{R^3}
	\right)
	\tanh
	\left(
	\frac{\xi_{mn} H_l}{R}
	\right),
\end{equation}

\noindent
where $m,n$ are the indices defining the sloshing mode, $\xi_{mn}$ is the $n^{\text{th}}$ zero of the first derivative of the $m^{\text{th}}$ order Bessel function, $\sigma$ is the surface tension and $\rho_l$ is the liquid's density \cite{Ibrahim2005}. The lowest wave mode that can be excited in these conditions is the first asymmetrical one $m=n=1$.
Investigations on several flights of the Arianespace launchers show that the $(1,1)$ mode is the primary excitation during the propelled phase \cite{Arndt2011,Montsarrat2017}. Therefore, this work focuses on excitations close to $\omega_{11}$.

Three different wave responses may be observed in these conditions: (1) planar waves, (2) chaotic, and (3) swirl sloshing \cite{Miles1984}. The planar wave regime is observed for low amplitudes and frequencies and is characterized by harmonic motion. The chaotic regime is observed for excitation frequencies near resonance conditions. In this regime, the sloshing dynamics quickly depart from a linear condition, leading to unpredictable interface dynamics.
Finally, the swirling regime is observed for excitation frequencies slightly above resonance. Swirling waves can be classified as stable if the rotational velocity remains constant or unstable if it periodically changes direction. 

The boundaries between these regimes have been analytically derived by Miles \cite{Miles1984}, and experimentally verified by Royon-Lebeaud \cite{Royon-Lebeaud2007}, who proposed the following relation:

\begin{equation}
	\label{eq:miles}
	\frac{A_0}{R}
	=
	\frac{1}{1.684}
	\left(
	\frac{\left(\Omega/\omega_{11}\right)^2-1}{\mathcal{B}_i}
	\right)^{3/2},
\end{equation}

\noindent
where $A_0$ and $\Omega$ were defined in Section \ref{sec:problem_description} and $\mathcal{B}_i$, is the frequency offset parameter that separates the sloshing regimes in the vicinity of $\omega_{11}$. These are $\mathcal{B}_2 = -0.36$, $\mathcal{B}_3 = -1.55$, and $\mathcal{B}_4 = 0.735$.
% There are actually 6 different beta parameters. These are bifurcations of the system of ODEs that explain how weakly non-linear sloshing waves evolve. 
% B6 almost overlaps with B4. B5 goes a bit after the planar/swirl regime. B1 does not only exist in low-damping conditions.

\begin{figure}[h]
	\centering
	\includegraphics[width=1\linewidth]{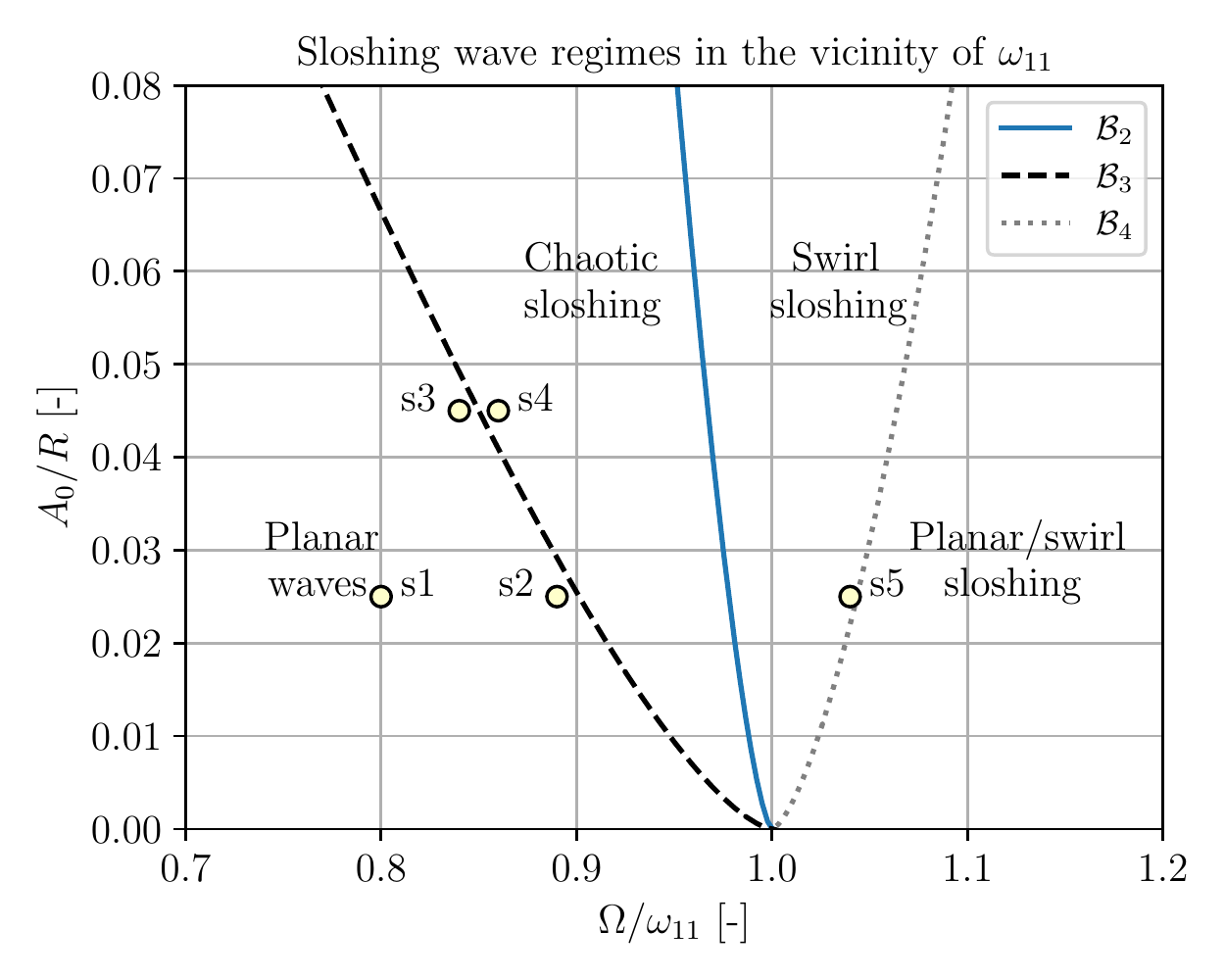}
	\caption{Sloshing regimes diagram from Miles' weakly nonlinear theory \cite{Miles1984}. The tested experimental cases are identified and labelled with numbers ranging from $1-5$. The excitation conditions are further detailed in Table \ref{tab:initial_conditions}.}
	\label{fig:sloshing_regimes}
\end{figure}

Figure \ref{fig:sloshing_regimes} shows Mile's diagram together with boundaries between regimes according to \eqref{eq:miles} and the six conditions investigated in this work. These are further detailed in Table \ref{tab:initial_conditions} and described in Section \ref{4p2} together with the experimental procedure. To complement the aforementioned kinematic description, the dynamics of the sloshing regime and its mixing efficiency are linked to the sloshing Reynolds number $\text{Re}_s$ proposed by Ludwig \& Dreyer \cite{Ludwig2013}, defined as:

\begin{equation}
	\label{eq:Re_s}
	\text{Re}_s
	=
	\frac{\Omega}{\omega_{11}}
	\left(
	\frac{b}{R}
	\right)^2
	\frac{\left( g R^3 \right)^{1/2}}{\nu}
	\sqrt{\xi_{11}},
\end{equation}

\noindent where $b$ is an estimate of the wave amplitude expected during the sloshing motion. We take $b/A_0 \approx 2[ (\Omega/\omega_{11})^2/(1- (\Omega/\omega_{11})^2 )  ]$ in the planar regime, $b \approx 0.54 R$ in the chaotic regime and $b \approx 0.7 R$ in the swirl sloshing regime, as indicated by \cite{Ludwig2013}. According to these authors, the critical Reynolds number for the mixing is $(\text{Re}_s)_c = 4.0 \cdot 10^3 \pm 20\%$. Below this value, sloshing does not significantly increase the heat or mass transfer in the tank.

%%%%%%%%%%%%%%%%%%%%%%%%%%%%%%%%%%%%%%%%%%%%%%%%%%%%%%%%
%%%%%%%%%%%%%%%%%%%%% METHODOLOGY %%%%%%%%%%%%%%%%%%%%%%
%%%%%%%%%%%%%%%%%%%%%%%%%%%%%%%%%%%%%%%%%%%%%%%%%%%%%%%%

\section{Methodology}
\label{sec:methodology}

\subsection{Experimental setup}
\label{sec:experimental_setup}

\begin{table*}[b]
	\centering
	\small
	\renewcommand{\arraystretch}{1.3}
	\caption{Location of the thermocouples (sorted by increasing height) with respect to a cylindrical coordinate system with origin centered in the tank's lower inner surface. The numbering of the thermocouples and orientation of the axes is illustrated in Figure \ref{fig:instrumentation_schematic}.}
	\begin{tabular}{c|ccccccccccccc}
		\hline
		Thermocouple no. {[}-{]} & 1    & 8    & 7    & 2    & 6    & 3    & 5    & 11  & 4    & 12   & 10    & 13   & 9     \\ \hline
		$z$ {[}mm{]}             & 9    & 19   & 34   & 39   & 44   & 46   & 49   & 50  & 52   & 64   & 69    & 84   & 94    \\
		$r$ {[}mm{]}             & 37.5 & 37.5 & 37.5 & 37.5 & 37.5 & 37.5 & 37.5 & 1.5 & 37.5 & 15.1 & 15.1  & 30.0 & 30.0  \\
		$\theta$ {[}$^\circ${]}  & 10   & 170  & 155  & 25   & 140  & 40   & 120  & 90  & 60   & 5.7  & 174.3 & 2.9  & 177.1 \\ \hline
	\end{tabular}
	\label{tab:tc_h_sort}
\end{table*}

\begin{figure}[h]
	\centering
	\includegraphics[width=0.97\linewidth]{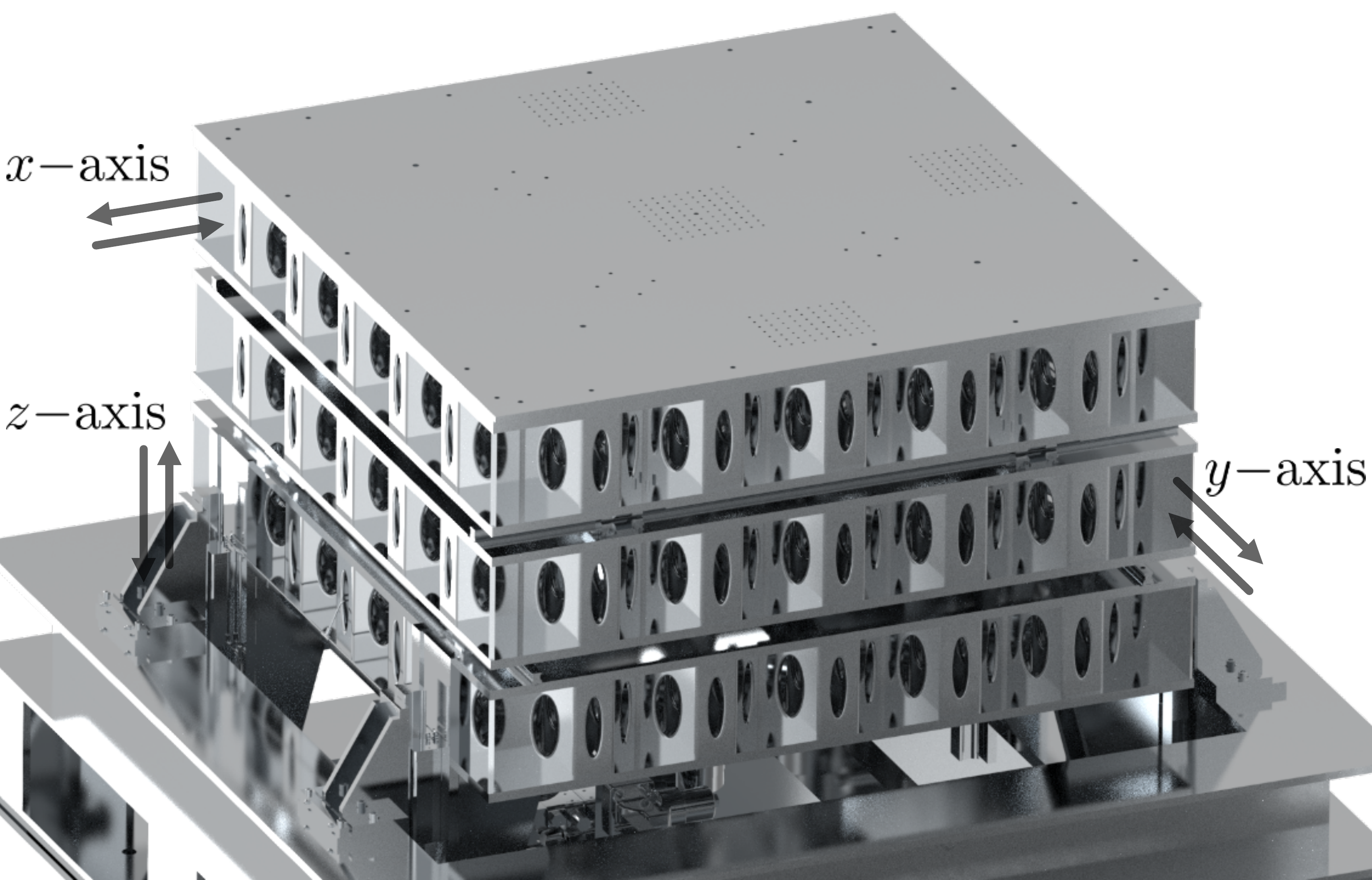}
	\caption{3D view of the three sliding modules of the SHAKESPEARE (SHaking Aparatus for Kinetic Experiments of Sloshing Projects with EArthquake Reproduction) shaking table facility.}
	\label{fig:shakespeare}
\end{figure}

The experimental campaign was carried out in the SHAKESPEARE (SHaking Aparatus for Kinetic Experiments of Sloshing Projects with EArthquake Reproduction) facility in the von Karman Institute. The table has dimensions of 1.5 m $\times$ 1.5 m and allows to mount different experimental setups. This facility is composed of three sliding modules, one for each axis, which can recreate complex three-dimensional excitation with controlled amplitudes and frequencies. A schematic of the table's three sliding modules is shown in Figure \ref{fig:shakespeare}. Since this work focuses on lateral sloshing, only excitation in the $x$ direction was considered. The displacement of this sliding module was measured with an Optical Displacement Sensor (ODS30) \cite{ODS_30}. 
This sensor measures displacements in the range of $28 - 32$ mm.

\begin{figure}[h]
	\centering
	\includegraphics[width=0.8\linewidth]{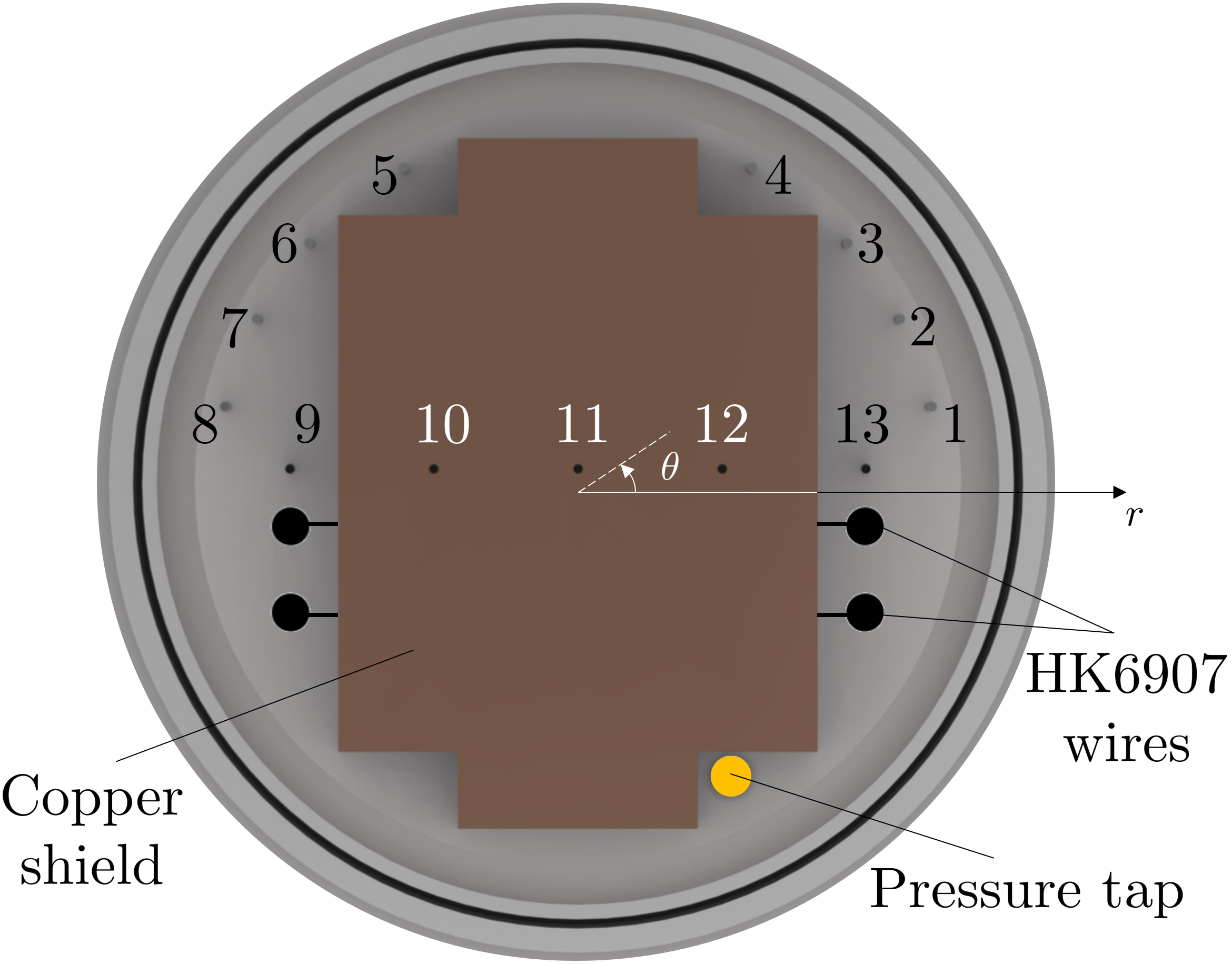}
	\caption{Schematic of the top view of the instrumented sloshing cell. Thirteen 0.25mm type K thermocouples, one static pressure tap and two thermofoil heating elements were mounted on the top-cover of the sloshing cell.}
	\label{fig:instrumentation_schematic}
\end{figure}

The sloshing cell used in this work is a quartz rectangular cuboid with dimensions 100 mm $\times$ 100 mm $\times$ 134 mm in which a cylindrical hole with diameter $D=80$ mm was drilled down to a depth of 124 mm. 
% The shape of the quartz cell was designed to withstand the harsh testing conditions, thermal stresses and heat cycles during the cryogenic experimental campaigns performed over the years in the von Karman Institute. Moreover, the optical access provided by the quartz material allows for the application of non-intrusive optical techniques to characterize the sloshing dynamics \cite{Simonini2018}.
The top cover of the cell consisted of a 12 mm acrylic glass plate with a cylindrical extrusion in the centre with a 12mm thickness and a diameter slightly smaller than 80 mm. This assembly was instrumented with thirteen 0.25 mm grounded mineral insulated type K thermocouples, a static pressure tap, and two heating elements.
Figure \ref{fig:instrumentation_schematic} shows a top view of the cell, illustrating the position of the pressure tap and the thermocouples in radial and angular coordinates. The thermocouples are numbered counter-clockwise, and their location is given in Table \ref{tab:tc_h_sort} in a cylindrical reference system centred on the bottom wall, at the origin of axes in Figure \ref{fig:problem_description}. Eight thermocouples were placed in one half of the cell's cross-section along its periphery at a diameter $D = 75$ mm, and five were located along the centerline. These thermocouples have a response time of $7.5$ ms when plunged into boiling water from air at 20 $^\circ$C, and their signals were acquired through two NI9212 8 Channel National Instruments cards with a sampling rate of 95 Hz per channel.

% The thermocouples were concentrated around $z=60$ mm since this was the height of the interface during the experimental campaign.

The top-mounted pressure tap was connected to a Validyne with an M38 diaphragm diameter and maximum admissible pressure of 0.55 bar. The Validyne CD15 carrier demodulator was used to convert the analogue pressure measurement into an output voltage between $\pm 10$ V. The response time of the pressure line was estimated to be around 156 ms.

Two Minco Polyimide Thermofoil HK6907 heaters \cite{HK6097} we\-re selected to impose the thermally stratified field during the initialization stage. The heating was applied from the top to reproduce an axial thermal stratification similar to \cite{Lacapere2009,Arndt2011,Ludwig2013}.
These heaters have dimensions of 25.4 mm $\times$ 50.8 mm $\times$ 1.14 mm, and minimum and maximum working temperatures of -32 and 100 $^\circ$C, respectively. Their maximum output power is 16.4 W. A 0.25 mm type K thermocouple was attached to one of these elements with thermal paste to monitor their temperature during the heating stage of the experiments. This thermocouple was then connected to a Tempatron PID330 controller to regulate the power provided by the TENMA 72-8700 \cite{Tenma_27_8700} power supply, ensuring the maximum limit of $100 ^\circ$C was never crossed.

\begin{figure}[H]
	\centering
	\begin{subfigure}[a]{0.9\linewidth}
		\includegraphics[width=\linewidth]{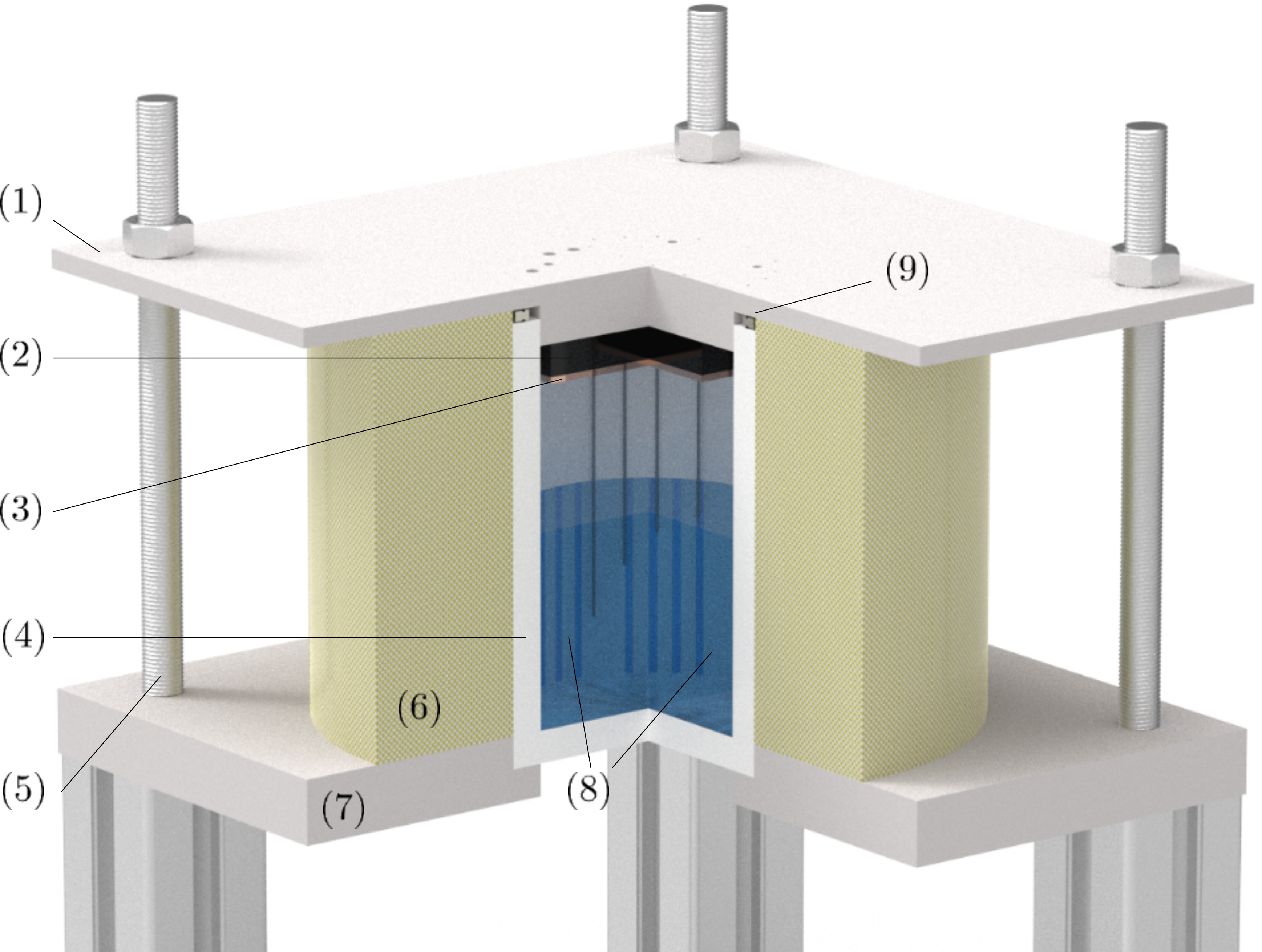}
		\caption{}
		\label{fig:facility_insulation}
	\end{subfigure}
	\hfill
	\begin{subfigure}[b]{0.9\linewidth}
		\includegraphics[width=\linewidth]{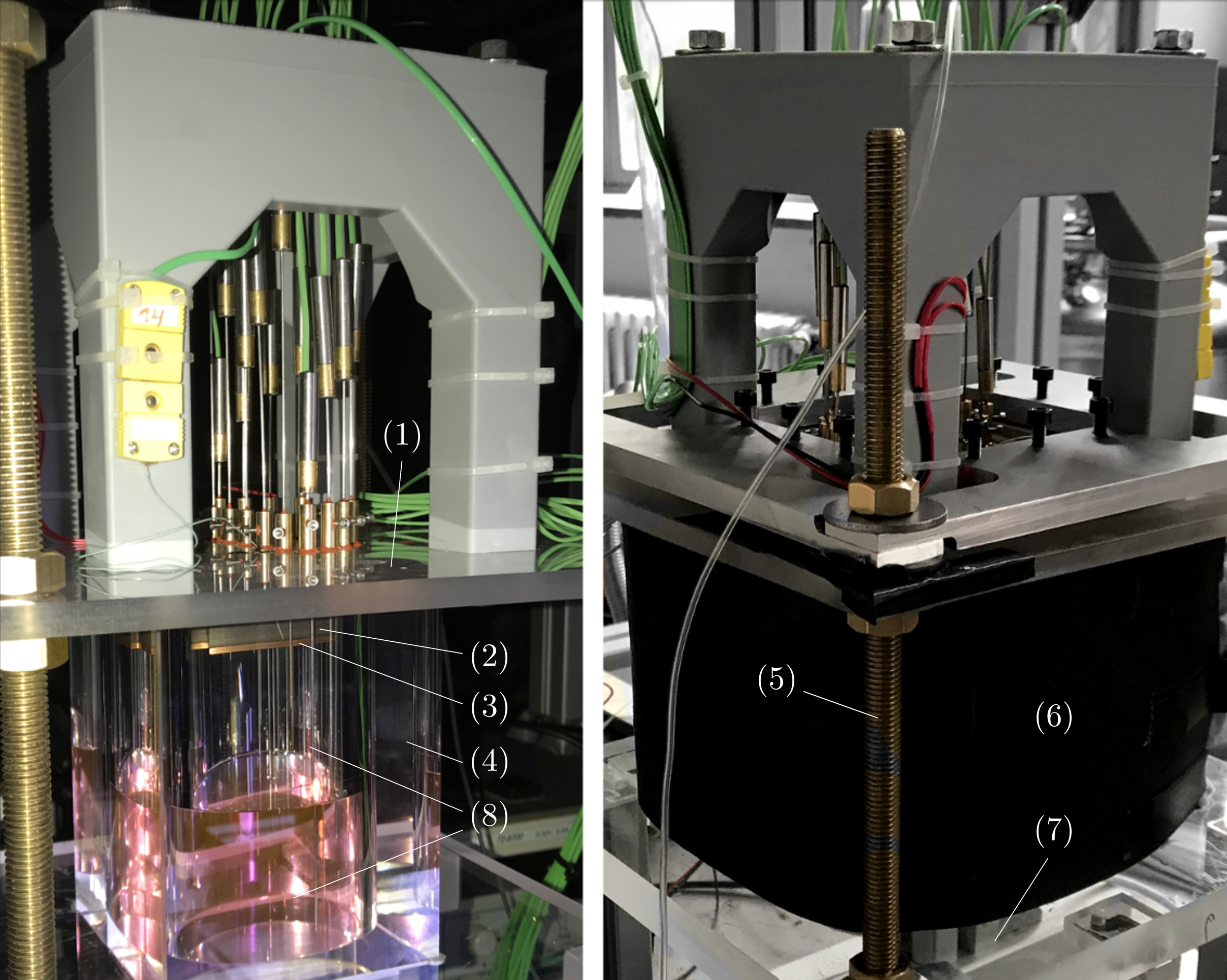}
		\caption{}
		\label{fig:facility_insulation_pictures}
	\end{subfigure}
	\caption{Instrumented experimental setup. a) Sectional view of the different elements which compose the experimental setup: (1) plexiglas top-cover, (2) bakelite insulating plate, (3) copper shield which houses both HK6907 Minco heating elements, (4) quartz tank, (5) steel threads that allow for screwing down the top-cover assembly, (6) Rockwool 850 insulating layer, (7) plexiglas bottom-cover, (8) temperature sensing assembly, (9) double-grooved aluminium interface plate with two O-rings. b) Pictures of the set-up without (left) and with (right) insulation.}
	\label{fig:facility_insulation_both}
\end{figure}

The facility was insulated with a 50 mm layer of Rockwool 850 material \cite{Rockwool_850}. This minimized the environmental effect and improved the experiments' repeatability. Figure \ref{fig:facility_insulation} shows a cut-out view of the experimental setup composed of the quartz tank, top-cover and bottom-cover assemblies, temperature sensing instrumentation, and insulation layer.

% \begin{figure}[h]
	% 	\centering
	% 	\includegraphics[width=1\linewidth]{figures/sec_4/setup_isometric_view.png}
	% 	\caption{Section view of the different elements which compose the experimental setup: (1) plexiglas top-cover, (2) bakelite insulating plate, (3) copper shield which houses both HK6907 Minco heating elements, (4) quartz tank, (5) steel threads that allow for screwing down the top-cover assembly, (6) Rockwool 850 insulating layer, (7) plexiglas bottom-cover, (8) temperature sensing assembly, (9) double-grooved aluminium interface plate with two O-rings.}
	% 	\label{fig:facility_insulation}
	% \end{figure}

The facility was equipped for high speed visualization (HSV) in back-lighting conditions and for Particle Image Velocimetry (PIV). Both were performed without insulation, which prevents optical access. Nevertheless, the importance of these tests was twofold, as further described in Section \ref{5p1}. The HSV allowed a qualitative confirmation of the expected sloshing regime while the PIV was performed to analyze the flow field in the liquid phase \emph{before} the sloshing event and thus ensure that each experiment started from the same (quiescent) conditions. For the PIV measurements, the liquid was seeded with fluorescent polymer microspheres with a density of 1300 kg/m$^3$ and diameter 1-5 $\mu$m.
The image acquisition was performed with a 12 bit Imager SX 4M camera \cite{Davis8} with a resolution of 2360 × 1776 pixels. The region of interest was illuminated by the Litron Bernoulli Laser \cite{LitronBernoulli}. This pulsed Nd:YAG laser has a maximum repetition rate of 15 Hz, an output power of 200 mJ per cavity, and a $3-7$ ns pulse duration at 532 nm. The laser beam was focused through the spherical lens and turned into a sheet with a cylindrical lens. This sheet was then deflected through a prism underneath the sloshing cell, thus illuminating its mid-plane from the bottom.

\subsection{Experimental procedure and test matrix}\label{4p2}

\begin{table*}[t]
	\centering
	\small
	\renewcommand{\arraystretch}{1.3}
	\caption{Summary of the initial thermodynamic conditions at the start of the pressurization and sloshing stages, and the investigated lateral sloshing excitation. Superscripts `$Q$' and `$T$' refer to the start of the constant heat-flux and temperature heating regimes, respectively. The superscript `$S$' is associated with the start of sloshing. Average gas and liquid temperatures $\tilde{T}_g$ and $\tilde{T}_L$, respectively, are presented at the start of the heating stage $t^{Q}$, and at the start of sloshing $t=0$. In this table, $A_0$ is the amplitude of the lateral motion, $R$ is the tank radius, $\Omega$ is the imposed angular frequency, $\omega_{11}$ is the natural frequency of the $(1,1)$ sloshing mode, $f_e$ is the excitation frequency in Hertz, and Re$_s$ is the sloshing-based Reynolds number \cite{Ludwig2013}.}
	\begin{tabular}{c|ccccccccccccc}
		\hline
		Case & $t^Q$ {[}s{]} & $t^T$ {[}s{]} & $p_{g}^{Q}$ {[}kPa{]} & $\tilde{T}_{l}^{Q}$ {[}K{]} & $\tilde{T}_{g}^{Q}$ {[}K{]} & $p_{g,0}^{S}$ {[}kPa{]} & $\tilde{T}_{l,0}^{S}$ {[}K{]} & $\tilde{T}_{g,0}^{S}$ {[}K{]} & $A_0/R$ & $\Omega/\omega_{11}$ & $A_0$ {[}mm{]} & $f_e$ {[}Hz{]} & Re$_s$ {[}-{]} \\ \hline
		s0    & $-$3602       & $-$3129       & 100.0                 & 291.5                       & 291.4                       & 103.5                   & 295.1                         & 313.9                         & -       & -                    & -              & -              & -              \\
		s1    & $-$3613       & $-$3179       & 99.4                  & 297.7                       & 296.9                       & 102.5                   & 298.4                         & 315.6                         & 0.025   & 0.80                 & 1.0            & 2.70           & 5.1e3          \\
		s2    & $-$3622       & $-$3198       & 100.0                 & 297.4                       & 296.9                       & 103.5                   & 298.7                         & 315.8                         & 0.025   & 0.89                 & 1.0            & 3.00           & 1.2e4          \\
		s3    & $-$3610       & $-$3178       & 99.3                  & 296.8                       & 296.8                       & 102.6                   & 298.3                         & 315.3                         & 0.045   & 0.84                 & 1.8            & 2.83           & 1.3e4          \\
		s4    & $-$3607       & $-$3167       & 99.9                  & 296.6                       & 296.7                       & 103.7                   & 298.7                         & 315.7                         & 0.045   & 0.86                 & 1.8            & 2.90           & 1.6e4          \\
		s5    & $-$3610       & $-$3185       & 99.8                  & 297.0                       & 296.6                       & 103.3                   & 298.4                         & 315.6                         & 0.025   & 1.04                 & 1.0            & 3.5            & 5.1e4          \\ \hline
	\end{tabular}
	\label{tab:initial_conditions}
\end{table*}

The experimental procedure and the test matrix considered in this work are described in this section. Each experiment consists of four phases, three of which are preparatory. In the first phase, the quartz 
cell is filled with $302 \pm 2$ ml of HFE-7200. This results in a liquid level of $60 \pm 0.4$ mm measured from the bottom of the cell. Then, the top cover is closed and screwed into position. Temperature-resistant grease is used to improve the connection between the O-rings, the aluminium interface plate, the quartz cell and the top cover. The setup remains closed for several hours until stable temperature and pressure readings are established inside the cell. This duration is long enough to allow the liquid's vapour to saturate the ullage gas.

Once this condition was met, the second phase begins with the acquisition of the pressure and temperature signals. During this second phase, the heating elements are switched on and produce a constant heat flux until the operating temperature of 100 $^\circ$C is reached. Then, the third phase begins with the PID controller actively enforcing a constant temperature on the heater surface. During this third phase, the system is warmed at a slower rate from the constant temperature surface at the top of the cell.

\begin{figure}[H]
	\centering
	\begin{subfigure}[a]{\linewidth}
		\includegraphics[width=0.9\linewidth]{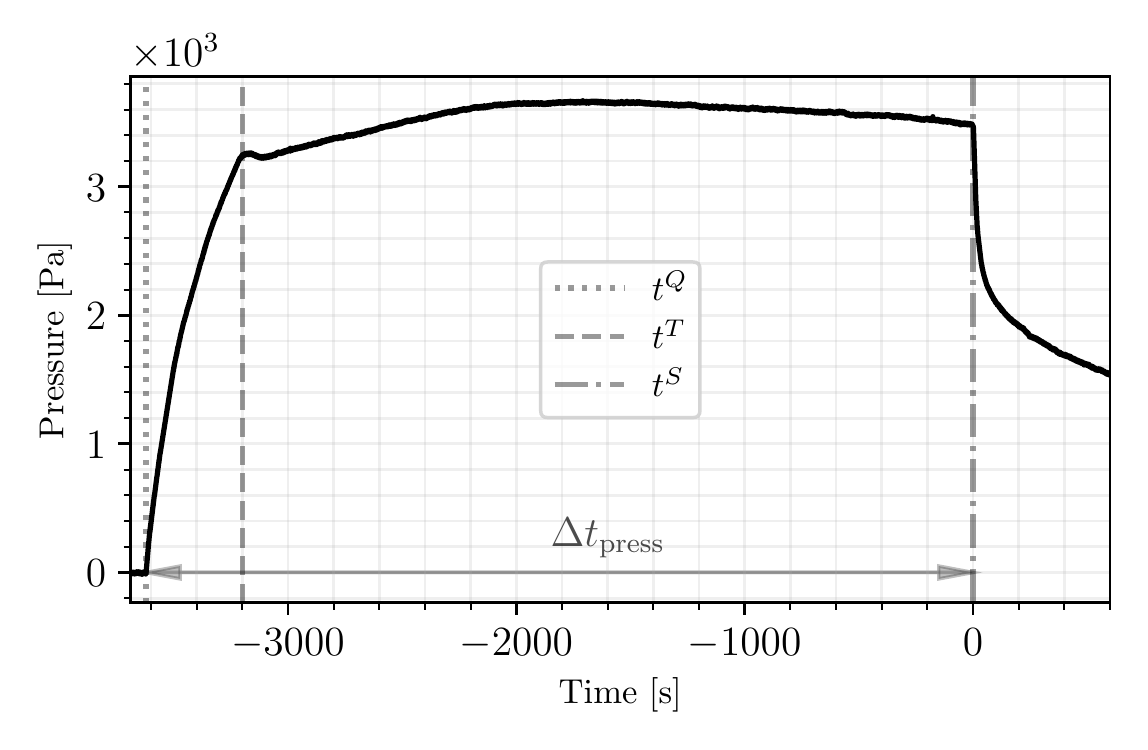}
		\caption{}
		\label{fig:sample_evolution_p}
	\end{subfigure}
	\hfill
	\begin{subfigure}[b]{\linewidth}
		\includegraphics[width=0.95\linewidth]{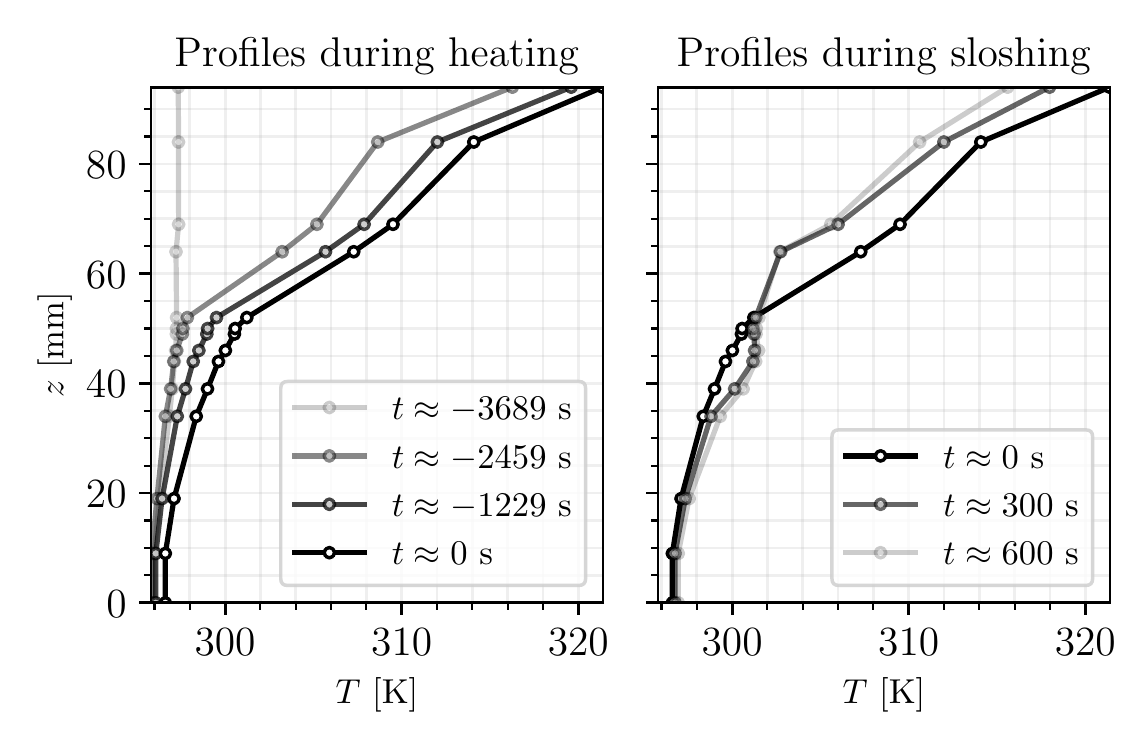}
		\caption{}
		\label{fig:sample_evolution_T}
	\end{subfigure}
	\caption{Thermodynamic evolution of the system during heating and sloshing in a sample experimental test case (Case s2 in Table \ref{tab:initial_conditions}). a) Gauge pressure evolution in the quartz cell and b) thermal profiles of the cell during heating (left) and sloshing (right). The system's heating starts at $t=t^Q$ with a constant heat flux until the heating elements' temperature reaches 100 $^\circ$C ($t=t^T$). It then proceeds at a constant temperature until $t=t^S=0$, when the heating is switched off, and the sloshing starts. The liquid interface is at $z=60\pm0.4$mm.}
	\label{fig:sample_evolution}
\end{figure}

The pressure evolution and temperature distribution in the cell were continuously measured throughout this initialization stage, which lasted one hour and generated an axial thermal stratification in the tank. Figure \ref{fig:sample_evolution} shows a typical reading for the pressure sensor in the ullage gas and the thermocouples during the second, third and fourth phases. The fourth phase consists of the actual sloshing experiment: the heater was switched off, and the lateral excitation was initiated from the controller screen of the SHAKESPEARE sloshing table. The time label $t=0$ is associated with the start of the sloshing, which is the focus of this work. Each excitation lasted 10 minutes, after which the data acquisition system was terminated, and the shaking table was turned off.

Figure \ref{fig:sample_evolution_p} shows the impact of the different phases on the pressure. The heating during the preparatory phase leads to an increase in the ullage pressure. This rise is initially steeper both because of the warm-up and the evaporation of the liquid. The temperature profiles, shown in Figure \ref{fig:sample_evolution_p}, clearly reveal the thermal stratification reached before sloshing. Sloshing mitigates the thermal stratification and reduces the temperature at the liquid interface by approximately 3 $^\circ$C in the illustrated example. This drastically reduces the saturation pressure in the ullage gas and induces condensation. The resulting pressure drop is visible in Figure \ref{fig:sample_evolution_p} for $t>0$s.

The initial thermodynamic conditions for each of the performed test cases are collected in Table \ref{tab:initial_conditions}. Case s0 was included as a reference condition to investigate the pressure and temperature evolution of the initialized system without sloshing. The positioning of these conditions in the Mile's diagram is shown in Figure \ref{fig:sloshing_regimes}.
%   Gauge pressure evolution in the quartz tank during a sample experimental test case. The heating of the system starts at $t^Q$ with a constant heat flux until the heating elements' temperature reaches 100 $^\circ$C. From this time, denoted as $t^T$, the system evolves with a constant temperature at the heater's surface. The total duration of the heating/pressurization stage is labelled as $\Delta t_\mathrm{press}$, and sloshing starts at $t = 0$ s. Figure obtained from `Case 2' (Table \ref{tab:initial_conditions}).
% Experimental thermal profiles measured during heating (left), and sloshing (right). The thermal stratification is disturbed due to the presence of sloshing, which mixes the warmer gas with the colder liquid in the vicinity of the interface. Figure obtained from `Case 2' (Table \ref{tab:initial_conditions}).
% Table \ref{tab:initial_conditions} summarizes the initial thermodynamic conditions for each of the performed test cases. Case s0 was included as a reference point to investigate the pressure and temperature evolution of the initialized system in the absence of sloshing. 
To compute the absolute pressure and spatially averaged temperatures required by the model, the measurement was complemented with a reading of the atmospheric pressure from the VKI's meteorological station while volume averaged temperatures were computed numerically at each time-instant through adaptive quadrature in Scipy \cite{SciPy} using the instantaneous readings in each thermocouple.

\subsection{Modeling of heat and mass transfer}
\label{sec:0d_model}

% The quasi-dimensional (0D) model described in this paper was based on the contents of Technical Note TN5000-10-05 from the cryogenics team of the von Karman Institute \cite{SloshII}. 

% The quasi-dimensional (0D) model is described in this section. As stated in Section \ref{sec:problem_description} the presented model relies on the assumption that the propellant tank has adiabatic walls and that the system is closed. Applying the principles of mass and energy conservation to this system yields:

The quasi-dimensional (0D) model used to determine the heat and mass transfer coefficients relies on the assumption that the propellant tank's wall are adiabatic and the system is closed. The mass and energy conservation thus give:

% \begin{figure}[h]
	% 	\centering
	% 	\includegraphics[width=0.6\linewidth]{figures/sec_2/0D_schematic.pdf}
	% 	\caption{Schematic of the 0D representation of the cryogenic propellant tank used in the upper stages of modern spacecraft. The effect of the walls was neglected in the current implementation of this model.}
	% 	\label{fig:0D_schematic}
	% \end{figure}

\begin{multicols}{2}
	\noindent
	\begin{equation}
		\label{eq:total_mass_conservation}
		\frac{dm_g}{dt} + \frac{dm_l}{dt} = 0
	\end{equation}
	\begin{equation}
		\label{eq:total_energy_conservation}
		\frac{dU_g}{dt} + \frac{dU_l}{dt} + \frac{dU_w}{dt}= 0,
	\end{equation}
\end{multicols}

\noindent where $m_g,m_l$ and $U_g,U_l$ are the masses and the internal energies of the ullage gas (subscript $g$) and the liquid (subscript $l$), and $U_w$ is the internal energy of the walls. The mass in the ullage gas is the sum of the inert gas ($m_a$, in our case air) and the liquid's vapour ($m_v$) masses, thus $m_g=m_a+m_v$. Assuming that the system is closed, i.e. $d m_a/dt=0$, the mass and the energy balances in each subsystem yield

\begin{equation}
	\label{eq:mass_ullage}
	\frac{dm_v}{dt} = -\frac{dm_l}{dt} = - \dot{m}_{i} 
\end{equation}

\begin{equation}
	\label{eq:energy_liquid}
	\frac{dU_l}{dt}
	=
	- \dot{Q}_{li}
	- \dot{m}_{li}  h_{l}^\text{s}
	+ \dot{Q}_{wl} 
	%\left(u_{L,\text{sat}} + \frac{p_v}{\rho_{L,\text{sat}}}\right)
\end{equation}

\begin{equation}
	\label{eq:energy_ullage}
	\frac{dU_g}{dt}
	=
	- \dot{Q}_{gi}
	+ \dot{m}_{i} h_{v}^\text{s}
	+ \dot{Q}_{wg}
	%\left(u_{v,\text{sat}} + \frac{p_v}{\rho_{v,\text{sat}}}\right)
\end{equation}

\begin{equation}
	\label{eq:energy_walls}
	\frac{dU_w}{dt}
	=
	- \dot{Q}_{wl}
	- \dot{Q}_{wg}
\end{equation}

\noindent
where $\dot{m}_{i}$ is the net mass flux from the gas to the liquid,  $h_{l}^\text{s}$ and $h_{v}^\text{s}$ are the saturated liquid and vapour enthalpies, and $\dot{Q}$ represents the heat fluxes linked to the heat transfer coefficients defined in Equations \eqref{eq:heat} and \eqref{eq:heat_tank}.

The interface temperature in \eqref{eq:heat} is assumed to be equal to the saturation temperature at the partial vapour pressure in the ullage gas, denoted as $p_v$. This can be estimated using the Clausius-Clapeyron relation \cite{Moran2011}:

\begin{equation}
	\label{eq:T_i}
	T_i = 
	\left( \frac{1}{T_\text{sat}^\text{ref}} 
	- 
	\frac{R_v}{\mathcal{L}_v}
	\ln\left(\frac{p_v}{p_{v,\text{sat}}^\text{ref}}\right)
	\right)^{-1}
	\,,
\end{equation}

\noindent
where $p_{v,\text{sat}} = 14.53$ kPa and $T_\text{sat}^\text{ref}=298.15$ K are the reference vapour pressure and corresponding saturation temperature and $\mathcal{L}_v=125.6$ kJ/kg is the latent heat of vaporization \cite{HFE7200_3M}. 
The net mass transfer due to the phase change is linked to the heat fluxes through the energy balance at the interface:

\begin{equation}
	\label{eq:mass_transfer}
	\dot{m}_{i} = - \frac{\dot{Q}_{gi} + \dot{Q}_{li}}{\mathcal{L}_v}\,.
\end{equation}

Equations \eqref{eq:total_mass_conservation}-\eqref{eq:energy_walls} can be combined into the rate of changes for the volume-averaged temperatures in the liquid, gas and tank walls:

\begin{equation}
	\label{eq:liquid_temperature}
	\frac{d\overline{T}_l}{dt} = 
	\frac{-\dot{m}_{i}
		\left(
		c_{p,l}\left(\overline{T}_l - T_i\right) - {p_v}/{\rho_l}
		\right)
		-\dot{Q}_{li}
		+\dot{Q}_{wl}
	}
	{m_l c_{p,l}}
\end{equation}

\begin{equation}
	\label{eq:ullage_temperature}
	\frac{d\overline{T}_g}{dt} = 
	\frac{\dot{m}_{i}
		\left(
		c_{p,v}\left(\overline{T}_g - T_i\right) - {p_v}/{\rho_v}
		\right)
		-\dot{Q}_{gi}
		+\dot{Q}_{wg}
	}{m_a c_{v,a} + m_v c_{v,v}}
\end{equation}

\begin{equation}
	\label{eq:wall_temperature}
	\frac{d\overline{T}_w}{dt} = 
	-
	\frac{
		\dot{Q}_{wl}
		+\dot{Q}_{wg}
	}{m_w c_{w}}
\end{equation}

\noindent
where $c_v$ and $c_p$ are the specific heat capacities at constant volume and constant pressure, respectively. Equations \eqref{eq:liquid_temperature} - \eqref{eq:wall_temperature} were derived considering that $dh \approx c_p dT$ and $du \approx c_v dT$ in the ullage, whereas $du \approx c_p dT$ for the liquid region as proposed in \cite{VanForeest2010}.

Given the lack of real gas equations of state for the HFE-7200 used in this work, the ideal gas law was used to model vapor's properties in the ullage. In a first approximation, this appears reasonable considering that the reduced pressure is in the range $p_R = p_v/p_c = 0.01 - 0.02$ and the reduced temperature is in the range $T_R = T/T_c = 0.60 - 0.65$, with critical conditions $T_c = 483.15$ K, and $p_c = 2.01$ MPa \cite{HFE7200_3M}. At such low reduced pressure, the compressibility factor $  Z = {p_v}/{\rho_v R_v T_g}$ can be expected to be close to unity for most common fluids \cite{Moran2011} and it this thus believed that the impact of this simplifying assumption is within the uncertainties of the inverse method. 

Therefore, the pressure evolution in the ullage can be determined by taking the time derivative of the ideal gas law and expanding it with the chain rule. This gives

\begin{equation}
	\label{eq:pressure_evolution}
	\frac{dp_g}{dt}
	=
	\frac{m_a R_a + m_v R_v}{V_g}
	\left(
	\frac{d\overline{T}_g}{dt}
	-
	\frac{dV_g}{dt}\frac{\overline{T}_g}{V_g}
	\right)
	+
	\frac{R_v \overline{T}_g}{V_g}
	\frac{dm_g}{dt}.
\end{equation}

Equation \eqref{eq:pressure_evolution} highlights the three fundamental mechanisms responsible for pressure variations in closed reservoirs. These are time variations in the (1) ullage gas temperature, (2) ullage volume, and (3) vapour mass. The rate-of-change of the ullage volume was derived from the volume variation on the liquid side. Since the liquid is treated as incompressible, its density is only a function of the temperature and thus 
% the chain rule on the time-derivative of the liquid volume yields

\begin{equation}
	\label{eq:V_L}
	\frac{dV_l}{dt}
	=
	- \frac{dV_g}{dt}
	=
	\frac{1}{\rho_l}\frac{dm_l}{dt}
	-
	\frac{m_l}{\rho_l^2}
	\left.\frac{\partial\rho_l}{\partial T}\right|_{p}
	\frac{d\overline{T}_l}{dt}
\end{equation}

\noindent
where $\left.\partial \rho_l/\partial T_l\right|_p$ is the liquid's isobaric compressibility.

% Same absolute pressure if we neglect hydrostatic and surface tension effects (but perhaps it's already obvious since we didn't include those in any of the aforementioned equations)

Under the assumption that the vapour and the inert in the ullage gas are at the same temperature $\overline{T}_g$ and the interface is always at saturation conditions (equation \eqref{eq:T_i}), the thermal evolution of the problem must be constrained to the condition that $\overline{T}_g > {T}_i$. However, the energy balance underlying \eqref{eq:liquid_temperature}-\eqref{eq:wall_temperature} does not guarantee this condition to be satisfied. To enforce this constraint, the system was complemented with the equations governing the evolution of the pressure and average temperatures of gas and liquid along the saturation line. These read:

\begin{equation}
	\label{eq:pg_dome}
	\frac{dp_g}{dt} 
	= 
	\left.\frac{\partial p_g}{\partial \rho}\right|_\text{sat}
	\left(
	\frac{1}{V_g}\frac{dm_v}{dt} - \frac{m_g}{V_g^2}\frac{dV_g}{dt}
	\right)
\end{equation}

\begin{equation}
	\label{eq:Tg_dome}
	\frac{d\overline{T}_g}{dt} 
	= 
	\left.\frac{\partial \overline{T}_g}{\partial \rho}\right|_\text{sat}
	\left(
	\frac{1}{V_g}\frac{dm_v}{dt} - \frac{m_g}{V_g^2}\frac{dV_g}{dt}
	\right)
\end{equation}

\begin{equation}
	\label{eq:Tl_dome}
	\frac{d\overline{T}_l}{dt} 
	= 
	\frac{Q_{wg} + Q_{wl} 
		- (m_a c_{v,a} + m_v c_{v,v})\frac{d\overline{T}_g}{dt}
		- c_{v,v}\overline{T}_g\frac{dm_v}{dt}
	}
	{m_l c_{p,l}}\,,
\end{equation}

\noindent
where $\left.\partial p_g/\partial \rho\right|_\text{sat}$ and $\left.\partial \overline{T}_g/\partial \rho\right|_\text{sat}$ were obtained by differentiating the Clasius-Clapeyron relation coupled with the ideal gas law.
These equations were solved in parallel to \eqref{eq:liquid_temperature}-\eqref{eq:wall_temperature} but activated only at the limiting case $\overline{T}_g = T_i$ to prevent $\overline{T}_g$ to drop below $T_i$.

Equations \eqref{eq:liquid_temperature}-\eqref{eq:Tl_dome} constitute a system of ordinary differential equations in which eight variables describe the state of the system. These are:

\begin{equation}
	\label{eq:state_x}
	\mathbf{x} =
	\begin{bmatrix}
		\overline{T}_g & 
		\overline{T}_l & 
		\overline{T}_w &
		m_v & 
		m_l &
		V_g &
		V_l &
		p_g
	\end{bmatrix}\,.
\end{equation}
The evolution of this system from a set of initial conditions $\mathbf{x}^0$ depends on the fluid properties and tank geometry, as well as four heat transfer coefficients $h_{li}, h_{gi}$, $h_{wl}, h_{wg}$ (see eq. \eqref{eq:heat}). 

Special treatment was given to the averaged wall temperature $\overline{T}_w$, because this was not measured during the experiments and cannot be linked to the other quantities unless its initial value $\overline{T}_w(t=0)$ is given. Therefore, the model predictions were run considering three scenarios for $\overline{T}_w(t=0)$. These are: (1) $\overline{T}_w(0)=\overline{T}_g(0)$, (2) $\overline{T}_w(0)=(\overline{T}_l(0)+\overline{T}_g(0))/2$ and (3) $\overline{T}_w(0)=\overline{T}_l(0)$. These cover the full range of possible wall temperatures and allow for analyzing the impact of this variable on the closure derivation.

Concerning the heat transfer coefficients $h_{li}$ and $h_{gi}$, experimental observation suggested that these could be modelled as decaying exponentials 

\begin{equation}
	\label{eq:h_gi_exp}  
	h_{gi}(t) = h^0_{gi}e^{-t/\tau} + h^\infty_{gi}
\end{equation}

\begin{equation}
	\label{eq:h_li_exp}
	h_{li}(t) = h^0_{li}e^{-t/\tau} + h^\infty_{li}\,,
\end{equation}

\noindent
which define $h^0_{gi}, h^0_{li}, h^\infty_{li}$, $h^\infty_{li}$ and $\tau$.
The time constant $\tau$ is related to the transient heat transfer due to the sloshing: within $t\ll\tau$, the heat exchange is strongly dominated by the sloshing-induced thermal de-stratification, while at $t\gg \tau$, the heat transfer is mostly due to conduction and hence less dependent on the sloshing regime. 
% Comment on h_{wg}
%\textbf{Finally, $h_{wg}$ was set to zero since...}
% The walls don't care because they have much higher thermal inertia
% Assume is mostly cooled through the interface
% Assume starting condition is such that the walls in contact with the gas are at similar temperature. The problem might be ill-condition if we considered this term (the optimization sure took a long time)

The model prediction in the proposed inverse method thus depends on seven parameters 

%are the coefficients that dictate the `steady' response of the system after the thermal gradients in the vicinity of the interface have stabilized. 
%The decaying constant $\tau$ was considered to be the same in both expressions since it was it was thought to be purely a function of the system’s excitation and initial thermodynamic conditions. Under this framework, the model must be supplied with the optimal set of coefficients:
%inverse method was implemented to determine the optimal set of coefficients:

\begin{equation}
	\mathbf{h} =
	\begin{bmatrix}
		h_{gi}^0 & 
		h_{li}^0 & 
		h_{gi}^\infty  & 
		h_{li}^\infty  & 
		\tau &
		h_{wg} &
		h_{wl}
	\end{bmatrix},
\end{equation}

\noindent which must be estimated from the available data. The global mass transfer coefficient defined in Equation \eqref{eq:mass} can be computed \textit{a posteriori} to evaluate the evaporation/condensation rates at the interface.

In summary, the thermodynamic model described in this section can be cast in the form of a parametric initial value problem:

\begin{equation}
	\systeme{
		\dot{\mathbf{x}} = \mathbf{f}{(\mathbf{x},t;\mathbf{h})},
		\mathbf{x}{(0)} = \mathbf{x}^0
	}
\end{equation}

\noindent
in which $\dot{\mathbf{x}}$ is the time-derivative of the state in \eqref{eq:state_x}, and $\mathbf{f}{(\mathbf{x},t;\mathbf{h})}$ collect the right-hand-side of the system of equations \eqref{eq:liquid_temperature}-\eqref{eq:Tl_dome}. In what follows, however, the problem set was further simplified by assuming that no heat exchange occurs between the gas and the wall (hence forcing $h_{wg}=0$). This simplification was motivated \textit{a-posteriori} because the optimization detailed in the following section systematically zeroed this term in all the investigated experimental conditions.

%======================================================%
%%%%%%%%%%%%%%%%%%%%%%%%%%%%%%%%%%%%%%%%%%%%%%%%%%%%%%%%
%%%%%%%%%%%%%%%%%%%% INVERSE METHOD %%%%%%%%%%%%%%%%%%%%
%%%%%%%%%%%%%%%%%%%%%%%%%%%%%%%%%%%%%%%%%%%%%%%%%%%%%%%%
%======================================================%

\subsection{Model-based inverse method}
\label{sec:model_based_inverse_method}

The inverse method consists in identifying the set of parameters $\mathbf{h}^*$ which minimizes the error between experimental data $\tilde{\mathbf{x}}_e{(\mathbf{t})}$ and the corresponding model outputs $\tilde{\mathbf{x}}(\mathbf{t})$. Here $\mathbf{t}\in\mathbb{R}^{n_t}$ denotes a set of randomly sampled points in time and the tilde denotes variable normalization. This scaling is necessary because the variables in $\mathbf{x}$ include different physical quantities with largely different numerical values and physical units. The scaling is performed by mean shifting and standardization (i.e. to unitary standard deviation).

The optimal parameters are thus those that minimize the objective function $\mathcal{J}{(\mathbf{x},\mathbf{h})}$ with respect to $\mathbf{h}$:

\begin{equation}
	\label{eq:optimization}
	\underset{\mathbf{h}}{\mathrm{argmin}} \ \ 
	\mathcal{J}{(\mathbf{x}(\mathbf{t});\mathbf{h})}
	= 
	||\tilde{\mathbf{x}}_e{(\mathbf{t})} - \tilde{\mathbf{x}}{(\mathbf{x}^0,t;\mathbf{h})} ||_2^2= ||\mathbf{R}(\mathbf{h})||^2_2,
\end{equation} 

\noindent
where $||\bullet||_2$ denotes the induced $l_2$ norm of a matrix. For a given set of times $\mathbf{t}\in\mathbb{R}{n_t}$, the residual are stored into a matrix $\mathbf{R}(\mathbf{h})\in\mathbb{R}^{8\times n_t}$.

The optimization was performed with the Basinhopping algorithm \cite{BasinHopping}. This alternates a quasi-Newton method with a random search for re-initialization. The random search is carried out using simulated annealing: a new candidate solution $\mathbf{h}^\text{new}$ is generated by perturbing the current guess, and this is accepted as a new solution with a piece-wise probability density of the form

\begin{equation}
	\mathcal{P}=
	\left\{\begin{matrix}
		1 & \mathrm{if} \ \ \ \mathcal{J}{(\mathbf{h^\text{new}})} < \mathcal{J}{(\mathbf{h^\text{old}})}
		\\[0.5em] 
		\exp{\left(
			-
			\frac{\mathcal{J}{(\mathbf{h^\text{new}})} - \mathcal{J}{(\mathbf{h^\text{old}})}}
			{\lambda}
			\right)} & \mathrm{if} \ \ \ \mathcal{J}{(\mathbf{h^\text{new}})} \geq \mathcal{J}{(\mathbf{h^\text{old}})}
	\end{matrix}\right.
\end{equation}

The parameter $\lambda$ plays the role of the temperature in simulated annealing, where it is usually made iteration-dependent. In this work, this was taken as $\lambda=0.01$. The new solution is then given as an initial guess for the quasi-Newton L-BFGS-B algorithm (see \cite{LBFGS}), which operates using finite differences for the gradient calculation. 
Once the stopping criterion for the BFGS is reached, the Basinhopping algorithm repeats the random search unless the optimization is terminated. The perturbation is made large enough to pick a new guess outside the basin of attraction of the identified local minima but small enough not to exceed the pre-defined boundaries of the parameter space. This was achieved by placing the new points halfway between the identified local minima. The termination occurs if the optimal solution does not change by more than $0.1\%$ in the last $500$ iterations.

The uncertainty in determining the optimal coefficients was estimated via a Monte Carlo method. Given that the measured signals contain noise, the input data in $\tilde{\mathbf{x}}_e{(\mathbf{t})}$ was randomly sampled from the available experiments such that only half of the points were used for training the model. In each of these sampling, the first entry in time is used as the initial condition for the numerical simulation, and the model calibration was performed $N=100$ times.

The initial conditions $\mathbf{x^0}$ supplied to the model were given as the starting values of the training set $\tilde{\mathbf{x}}_e(0)$.
Therefore, the model calibration was performed through $N = 100$ optimizations by randomly selecting different training sets and initial conditions each time. This resulted in a matrix of optimal solutions $\mathbf{H}^*\in \mathbb{R}^{100\times 5}$ in which the i-th row collects the optimized parameters at the i-th run. The variability in this matrix reflects the uncertainties in the model's training; thus, the standard deviation in each column is associated with the uncertainty in each parameter. Finally, these uncertainties were propagated to the numerical predictions of $p_g(t)$, $\overline{T}_g(t)$ and $\overline{T}_l(t)$ by evaluating the model for each set of coefficients in $\mathbf{H}^*$. 

In each of these evaluations, the initial conditions were perturbed with Gaussian distributions with mean and standard deviation taken from the experimental data of $p_g$, $\overline{T}_g$ and $\overline{T}_l$ in the quasi-stationary conditions before the sloshing.

%==============================================================%
%%%%%%%%%%%%%%%%%%%%%%%%%%%%%%%%%%%%%%%%%%%%%%%%%%%%%%%%%%%%%%%%
%%%%%%%%%%%%%%%%%%%%% RESULTS & DISCUSSION %%%%%%%%%%%%%%%%%%%%%
%%%%%%%%%%%%%%%%%%%%%%%%%%%%%%%%%%%%%%%%%%%%%%%%%%%%%%%%%%%%%%%%
%==============================================================%

\section{Results and discussion}
\label{sec:results}

This section is dedicated to presenting and discussing the main results of this research campaign. Section \ref{5p1} experimentally validates the assumptions made in Section \ref{sec:problem_description} on the boundary and initial conditions for the sloshing experiment.
Then, Section \ref{sec:5p2} describes the wave dynamics of the tested sloshing excitation by comparing the transient interface displacement of cases s1 (subcritical planar waves), s3 (supercritical planar waves) and s5 (swirl waves). The consequent sloshing-induced pressure drop and thermal mixing are documented in Section \ref{sec:p_T_exp}. The impact of different excitations is assessed, and the results are compared to reference case s0 in quiescent conditions. Finally, Section \ref{sec:model_calibration} reports on the model calibration through the inverse method described in Section \ref{sec:model_based_inverse_method}.

\subsection{Verification of initial and boundary conditions}\label{5p1}

% In this section, we aim to investigate the thermal boundary conditions of the system with and without the insulating layer, and to measure the initial velocity conditions at $t = 0$ s.

A preliminary experimental campaign was carried out to assess the adiabaticity of the cell. These experiments analyzed the temperature profile of the empty cell (i.e. filled with air) after one heating cycle (i.e., heating up the cell for one hour and allowing it to cool down passively for another hour). The tests were repeated with the insulated and the non-insulated facilities (see Section \ref{fig:facility_insulation_pictures}).

\begin{figure}[H]
	\centering
	\includegraphics[width=0.99\linewidth]{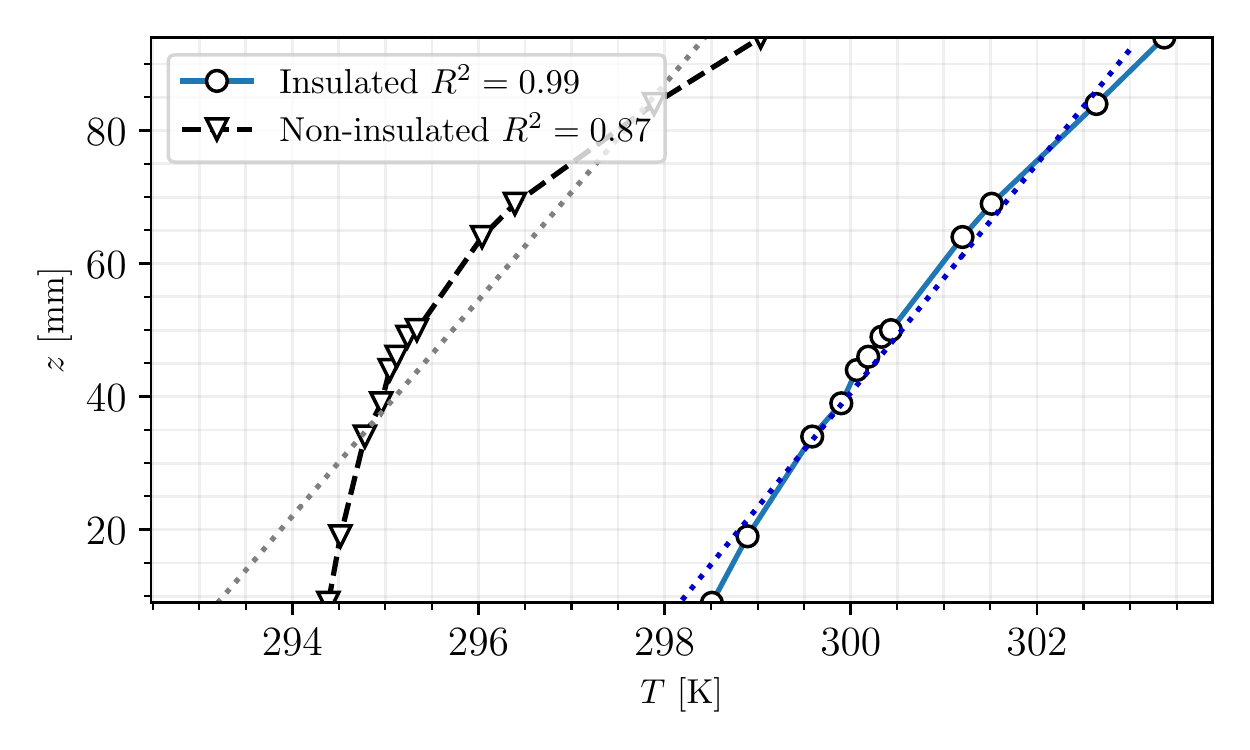}
	\caption{Thermal profiles in the non-insulated (black) and insulated (blue) setups filled only with air at $t=3600$ s. The heating procedure was initiated at $t\approx-3600$ s, and the heaters were turned off at $t=0$ s. The dotted lines are the optimal linear fit of each profile.}
	\label{fig:insulation_thermal_profiles_gas}
\end{figure}

During these tests, the thermal profile was analyzed with the available thermocouples (cf. Section \ref{sec:experimental_setup} and Table \ref{tab:tc_h_sort}). Under the assumption of fully adiabatic and axisymmetric conditions, and in the absence of buoyancy-driven flows, the conduction-dominated thermal field in the tank should lead to a linear temperature profile along its $z$ axis once steady-state conditions are reached.
The results of this analysis are reported in Figure \ref{fig:insulation_thermal_profiles_gas} for both the non-insulated and the insulated facilities. The linearity of the temperature profile proves that the assumption of adiabatic conditions is reasonable for the case of the insulated facility, which is the one analyzed in the following subsections.

A second campaign was also carried out, with the aim of characterizing the velocity field in the liquid during the heating phase. In this phase, before sloshing, the liquid should be quiescent to ensure repeatability. However, the temperature gradients in the cell inevitably produce buoyant flows that could undermine this assumption. Therefore, Particle Image Velocimetry (PIV) was used to analyze the potential impact of buoyancy in setting the initial conditions for the sloshing experiments. 
% The PIV was carried out using Fluorescent Polymer microspheres FM with a density of 1300 kg/m$^3$ in the non-insulated cell because of the need for optical access. 
Although the thermal profiles are different in the absence of the insulation, Figure \ref{fig:insulation_thermal_profiles_gas} shows that the $\Delta T$ between the top and the bottom surfaces of the cell are comparable. Therefore, it is believed that the PIV analysis in non-insulated conditions is also representative for the insulated setup.

\begin{figure}[H]
	\centering
	\includegraphics[width=0.99\linewidth]{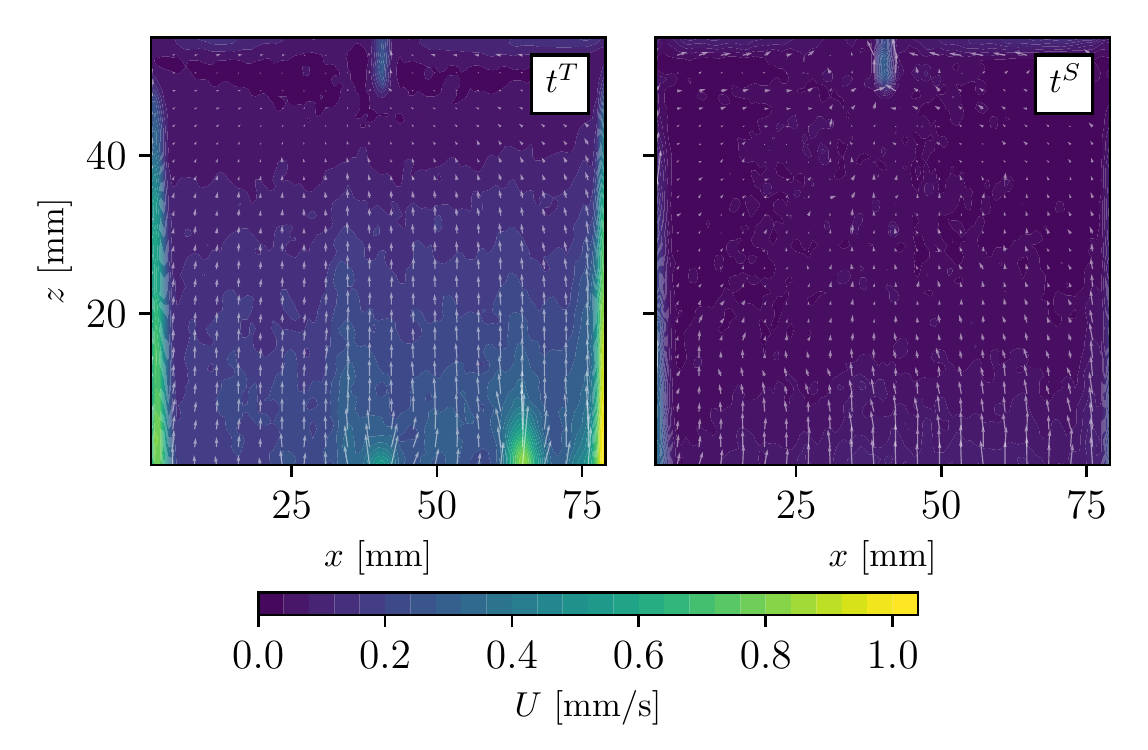}
	\caption{Velocity profiles captured at the start of the constant-temperature heating stage $t=t^T$ (left), and right before the start of sloshing $t=t^S$ (right).}
	\label{fig:PIV_U_fields}
\end{figure}

The cell was filled up to $H=60$mm, and the velocity field was captured at the start of the constant temperature phase (i.e. $t\in [t^T, t^S]$, see Figure \ref{fig:sample_evolution} and Section \ref{4p2}. The PIV was sampled at $10$ Hz, and time-averaged velocity fields were captured every 500 images. The first (at $t\approx t^T$) and the last (at $t\approx t^S$) time-averaged velocity fields are shown in Figure \ref{fig:PIV_U_fields}.

\begin{figure*}[h]
	\centering
	\begin{subfigure}[a]{\linewidth}
		\centering
		\includegraphics[width=0.90\linewidth]{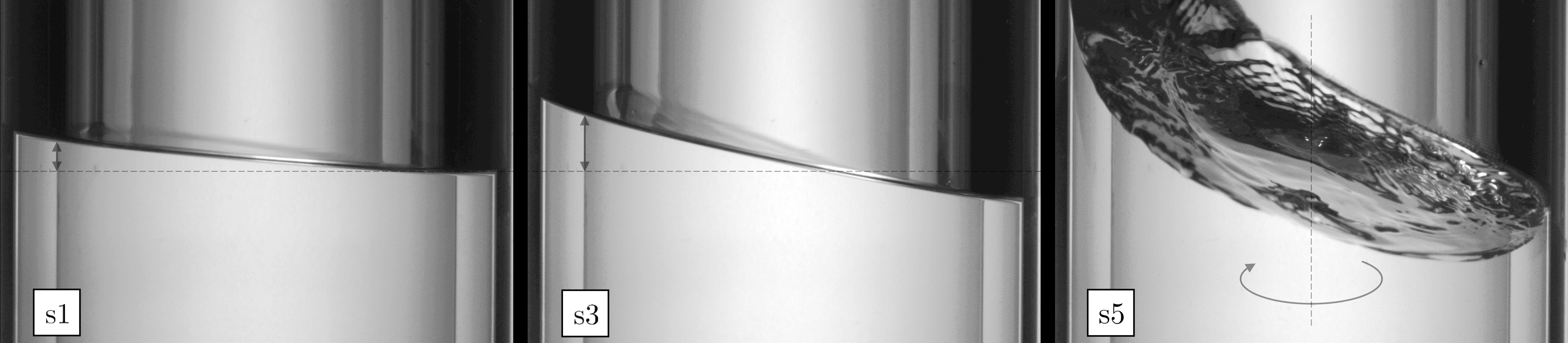}
		\caption{}
		\label{fig:sloshing_regimes_photos}
	\end{subfigure}
	\hfill
	\begin{subfigure}[b]{\linewidth}
		\centering
		\includegraphics[width=0.99\linewidth]{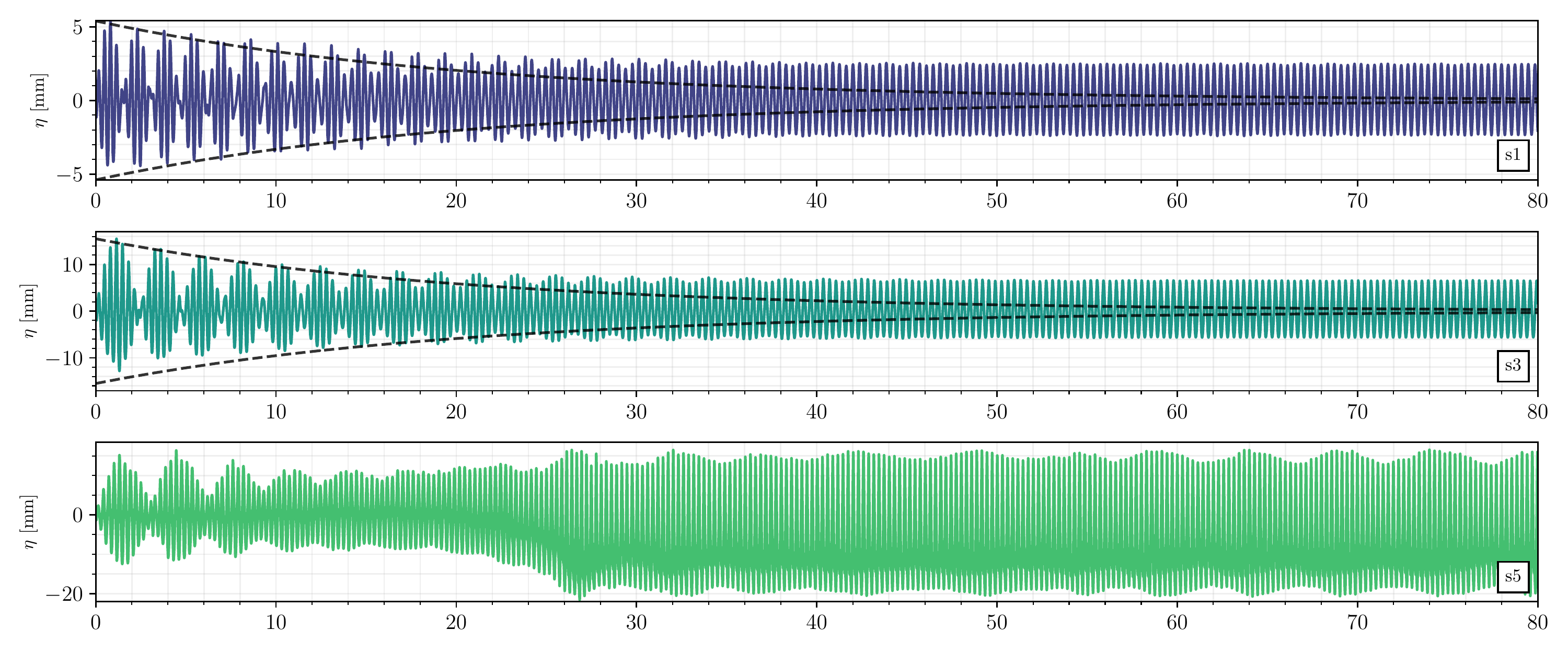}
		\caption{}
		\label{fig:interface_displacement}
	\end{subfigure}
	\caption{a) Snapshots of the sloshing motion triggered by the lateral excitation in cases s1, s3 and s5. Cases s1 and s3 are characterized by lateral interface displacement, which remains approximately flat and follows a harmonic motion. On the other hand, case 6 displays a stable rotational velocity which induced a high-amplitude swirl wave. b) Interface displacement measured at a point located in the mid-plane of the quartz tank at a radial distance $R/4$ from the side wall during the first 80 seconds of sloshing.}
	\label{fig:interface_signal_and_photos}
\end{figure*}

As shown in Figure \ref{fig:PIV_U_fields} on the left, upward convective currents are observed at $t\approx t^T$, with velocities of the order of $\approx 1$mm/s. However, these decrease during the course of the heating stage and nearly vanish at $t=t^S$ as shown in Figure \ref{fig:PIV_U_fields} on the right. In conclusion, these results show that the impact of buoyancy-driven currents on the liquid is negligible, and the assumption of quiescent initial conditions for the sloshing phase is reasonable.

\subsection{Lateral sloshing dynamics}
\label{sec:5p2}

We here report on the response of the liquid interface for the experimental conditions detailed in Table \ref{tab:initial_conditions}. The observed responses were in line with Mile's regime classification (Figure \ref{fig:sloshing_regimes}) except for test case s4, which was expected to be in the chaotic regime but produced a planar wave response like cases s2 and s3. This discrepancy might be explained by the proximity of the point to the boundaries between these two regimes (see Figure \ref{fig:sloshing_regimes}). However, as conditions s3 are equally close to this boundary but on the planar wave regime side, it appears that the frequency offset parameter $\mathcal{B}_3$ might be slightly under-predicted by the theory.

Concerning the test cases in the planar waves regimes, case s1 is in subcritical conditions, Re$_s$ $<$ (Re$_s$)$_c$ in \eqref{eq:Re_s}, while cases s2, s3, s4 are in supercritical conditions, Re$_s$ $>$ (Re$_s$)$_c$ in \eqref{eq:Re_s}. Finally, case s5 is in the swirl wave regime as expected.
% Ludwig2013 actually has a swirl case. They don't show the temp/pressure evolutions but it's in the experimental matrix, and the results of that case are summarized in a table
% While various authors have reported on the thermodynamic response of the ullage gas during planar sloshing \cite{VanForeest2010,Arndt2011}, to the authors' knowledge this is the first documented experiment in swirling conditions.
The lack of points in the chaotic regime in the reported campaign stems from the difficulties in performing the measurements in this regime. Several conditions in chaotic sloshing conditions were reproduced during the campaign, but the violent response of the interface resulted in severe wetting/de-wetting of thermocouples and pressure taps, which led to unreliable measurements. 

Figure \ref{fig:sloshing_regimes_photos} illustrates a snapshot of the liquid interface in the three investigated conditions, namely planar subcritical sloshing (s1), planar supercritical sloshing (s2) and swirl sloshing (s5). For each of these, the figures in the bottom reports on the time evolution of the vertical displacement $\eta(t)$ at a point located in the mid-plane of the tank at a distance $R/4$ away from the lateral wall. This displacement is computed with respect to the original height of the interface at $t=0$. This measurement was obtained via image processing of the video sequence, by identifying regions of sharp changes in the image's intensity along each `pixel-column'. The edge detection approach is similar to the approaches in \cite{Alessia,Miguel_I}.

As expected, the interface response for each of the sloshing modes in the planar test cases is the sum of the harmonic response to the perturbation and exponentially vanishing transients of the form $\exp(-\gamma_{n,j} \omega_{n,j} t)$. The decay rates $\gamma_{n,j}$ of the transients are linked to the viscous damping and can be computed analytically using the linear theory in \cite{Ibrahim2005}, in which the flow velocity is assumed to be the sum of a potential contribution and a rotational one due to viscosity.

Focusing on the first mode $(n,j)=(1,1)$, by far the most energetic, the decay rate $\gamma_{1,1}$ is derived in \cite{Ibrahim2005} and reads

\begin{equation}
	\label{eq:damping_rate}
	\gamma_{1,1} = \frac{\delta_{1,1}}{\sqrt{4\pi^2 + \delta_{1,1}^2}}
\end{equation}

\noindent
where the logarithmic decrement, $\delta_{1,1}$, is 

\begin{equation}
	\label{eq:log_decrement}
	\delta_{1,1} =
	% 3.52
	2 \pi K
	\frac{\nu^{\frac{1}{2}}\left( 1 + 2\left(1-\frac{h}{R} \right)\right)}
	{R^{\frac{3}{4}}g^{\frac{1}{4}}\tanh^{\frac{1}{4}}\left(\xi_{1,1}\frac{h}{R}\right)\sinh\left(2\xi_{1,1}\frac{h}{R}\right)}
\end{equation}

\noindent
for upright cylinders without baffles, where $K\approx0.59$ is a parameter that accounts for the geometry of the tank \cite{Arndt2011}. This derivation, also verified experimentally in \cite{Stephens1962}, considers that the frictional losses between the oscillating flow field and the solid walls are limited to the Stokes boundary layer \cite{Miles1956}.

The exponentially decaying envelope of the transient response using \eqref{eq:damping_rate} is shown with dashed lines in Figure \ref{fig:interface_displacement} for cases s1 and s3. As the transient vanishes, the interface motion tends towards its harmonic response.
The agreement between experimental data and theory is remarkable. Moreover, the interface's response is also characterized by the `beating phenomena' during the transient, as the excitation frequency ($2.7$ Hz in s1 and $2.83$Hz in s3) is close to the natural frequency ($3.375$ Hz). This results in the periodic modulation at $f_b=0.67$ Hz for s1 and $f_b=0.54$ Hz for s3. Finally, concerning the swirling case (case s5), the linear theory illustrated for the previous test cases does not hold. The sloshing initially begins with lateral waves (at the associated beating) but gradually gains a rotational component until a stable swirl wave is produced at $t>30$s. The impact of these sloshing excitation on the thermodynamic state of the system is described in the following section.

\subsection{Pressure and temperature transducer measurements}
\label{sec:p_T_exp}

The data acquired from the instrumentation mounted on the top cover of the sloshing cell is analyzed in this section. The measurements were non-dimensionalized to better compare the different cases. The non-dimensional temperature on the gas and liquid side are defined, respectively, as:

\begin{equation}
	\label{eq:theta_g}
	\Theta_g(t) = \frac{T(t) - T_\text{sat,0}}{T_\text{t,0} - T_\text{sat,0}}\,,
\end{equation}

\begin{equation}
	\label{eq:theta_l}
	\Theta_l(t) = \frac{T(t) - T_\text{b,0}}{T_\text{sat,0} - T_\text{b,0}}\,,
\end{equation}

% Excellent synthesis, bravo Pedro.

\noindent
where $T_\text{t,0}$ is the temperature measured at $t = 0$ by the thermocouple closest to the top of the tank, $T_\text{b,0}$ is the temperature measured the same instant by the thermocouple closest to the bottom, and $T_\text{sat,0}$ is the corresponding saturation temperature.

The non-dimensional pressure is defined as the ratio between the instantaneous gauge pressure in the cell $\Delta p_g(t)$ and its initial value $\Delta p_{g,0}$: 

\begin{equation}
	\label{eq:delta_p_g}
	\Delta \hat{p}_g = \frac{\Delta p_g (t)}{\Delta p_{g,0}}.
\end{equation}

We focus on the gauge pressure instead of the absolute pressure because the objective was to assess the pressure change in the ullage due to sloshing. 

\begin{figure}[H]
	\centering
	\includegraphics[width=0.99\linewidth]{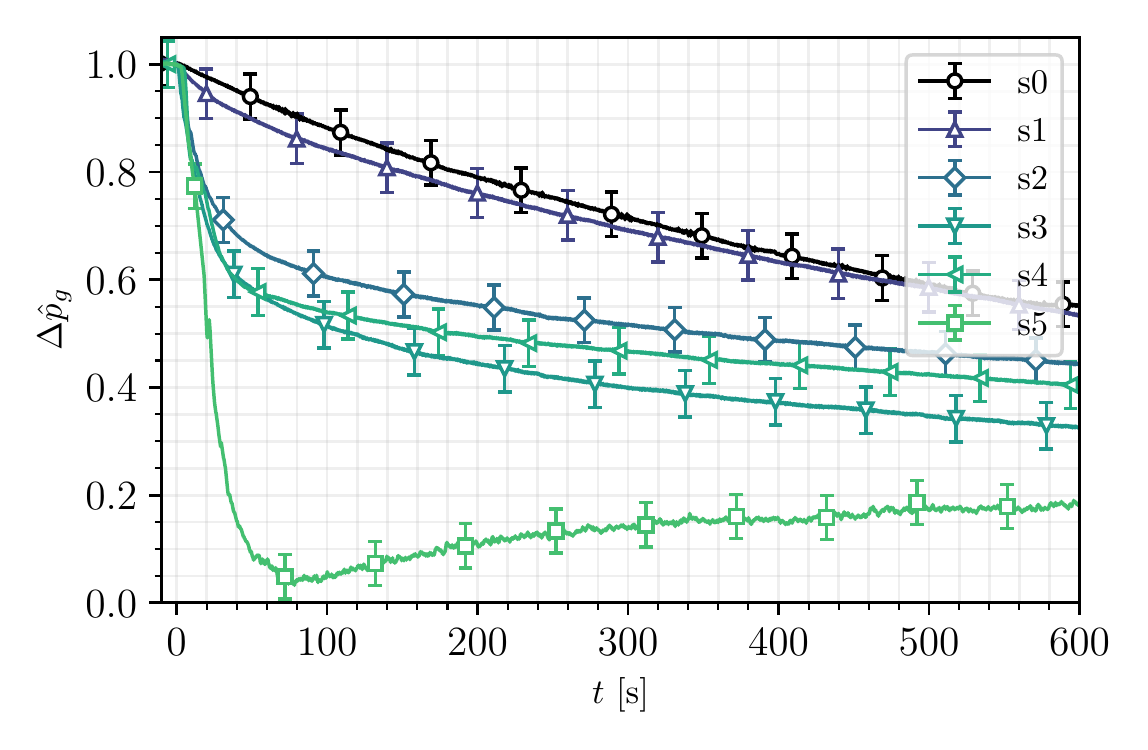}
	\caption{Normalized gauge-pressure in the quartz for experimental test cases s0-s5 after the pressurization/heating stage was concluded $t > 0$ s. The excitation of cases s1-s5 started at $t = 0$ s and lasted for 10 minutes. The legend underneath the figures indicates the height of each sensor with respect to the bottom of the tank.}
	\label{fig:pg_all_exp}
\end{figure}

\begin{figure*}[h]
	\centering
	\begin{subfigure}[a]{\linewidth}
		\centering
		\includegraphics[width=0.99\linewidth]{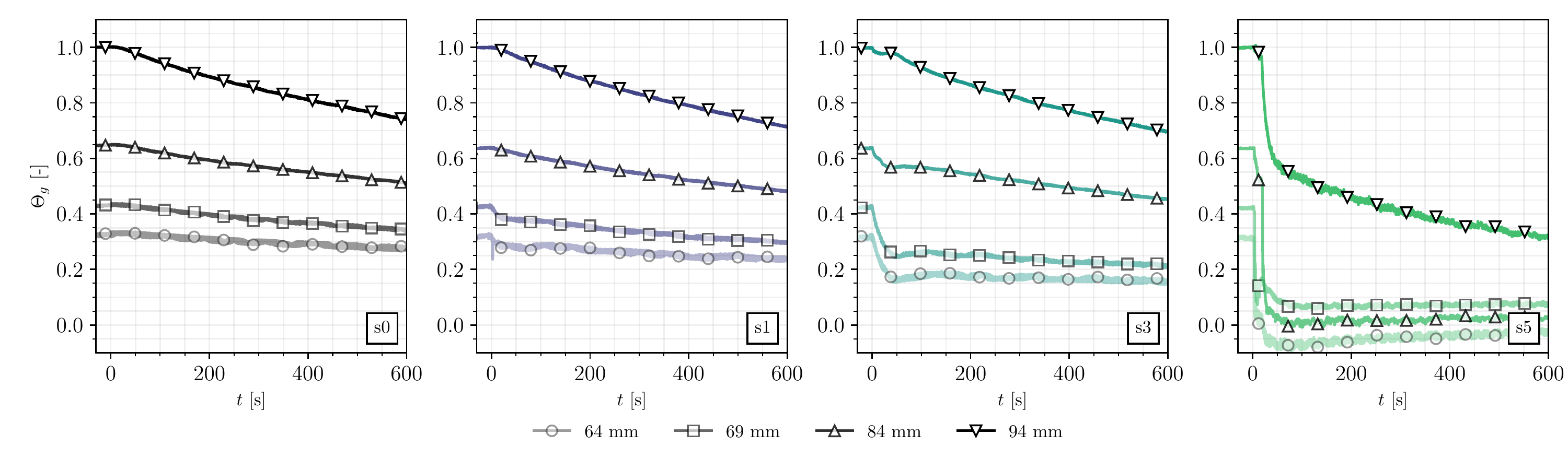}
		\caption{}
		\label{fig:theta_g_exp}
	\end{subfigure}
	\hfill
	\begin{subfigure}[b]{\linewidth}
		\centering
		\includegraphics[width=0.99\linewidth]{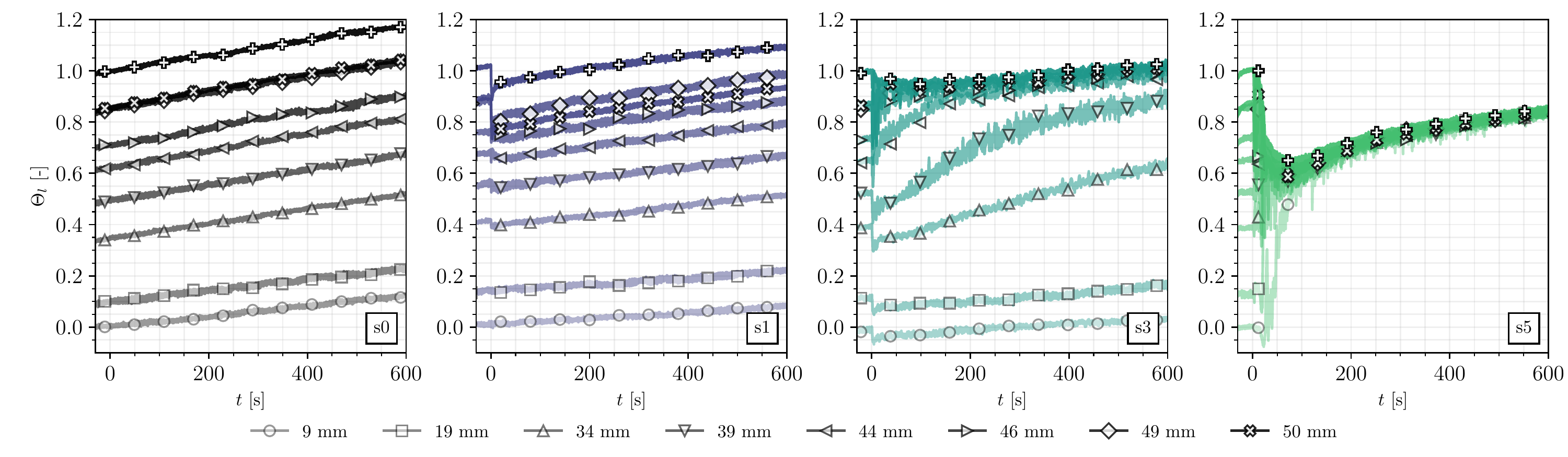}
		\caption{}
		\label{fig:theta_l_exp}
	\end{subfigure}
	\caption{Non-dimensional ullage (a) and liquid (b) temperatures of cases s0, s1, s3, and s5, as a function of time, measured by the thermocouples immersed in the liquid phase after the pressurization/heating stage was concluded $t > 0$ s.}
	\label{fig:theta}
\end{figure*}

Figure \ref{fig:pg_all_exp} shows the normalized gauge pressure evolution in all test cases after the system pressurization (phase three in Section \ref{4p2}) is concluded and the heating is turned off, that is $t \in [0, 10]$ min. This figure should be analyzed together with Figure \ref{fig:theta}, which shows the temperature evolution in the five cases, as measured by the thermocouples in the ullage volume (top row of plots) and the ones immersed in the liquid (bottom row). The height of the thermocouple location (see Table \ref{tab:tc_h_sort}) is recalled at the bottom of each line of plots.

We begin with case s0, which is the baseline test without sloshing. During the 10 minutes of observation, the gauge pressure decreased by roughly 45\%.
This is due to the cooling of the ullage to the lateral walls and the liquid under it, as visible from Fig \ref{fig:theta_l_exp}. Noteworthy is the continuous warm-up of the liquid due to the thermal energy stored in the thick quartz wall of the tank.

Moving to case s1, which is in planar sloshing conditions, its pressure evolution in Figure \ref{fig:pg_all_exp} is nearly identical to the baseline in quiescent conditions. The analysis of the temperature evolution reveals a moderate heat exchange between liquid and gas. On the gas side, this is noticeable only in the thermocouple closest to the interface (at approximately $4$ mm) near the interface. On the liquid side, a sloshing-induced temperature drop is observed down to the second thermocouple (about 40mm underneath the interface). However, soon after the sloshing event, the temperature evolution restores the linearly increasing trend observed for case s0. The moderate impact of sloshing on the pressure evolution in this test case is compatible with its low sloshing-based Reynolds number (see \eqref{eq:Re_s}), which is roughly 5.1e3 and thus within the $\pm 20 \%$ range of the critical value $(\text{Re}_s)_c$ below which no significant heat or mass transfer is expected \cite{Ludwig2013}.

Cases s2, s3, and s4 exhibited similar pressure and temperature evolutions. According to their sloshing-based Reynolds numbers, which were one order of magnitude higher than case s1, it was expected that these experiments should produce increasingly larger sloshing-induced pressure drops (i.e., s2 $<$ s3 $<$ s4). However, case s4 produced a pressure evolution higher than s2, but lower than s3. This might be explained by slightly higher initial temperatures in the quartz walls and the liquid bulk (Table \ref{tab:initial_conditions}). Nevertheless, the normalized pressure evolutions are within the uncertainty bars of each other, which was expected given that the sloshing-based Reynolds number was in the same order of magnitude across all three tests (Table \ref{tab:initial_conditions}). Because of this similarity, only the temperature distribution measured in case s3 is shown in Figure \ref{fig:theta_l_exp}. These cases are characterized by a sharp drop in the ullage pressure at the sloshing onset, followed by a strong thermal mixing in the vicinity of the interface. The sloshing event does not influence the temperature at the bottom of the tank (at about $51$ mm from the interface).

Finally, case s5, in swirling conditions, results in a significant pressure drop and thermal destratification. The gauge pressure (see Figure \ref{fig:pg_all_exp}) reduces by $90\%$ within the first 70 seconds, and the temperature in the liquid (see Figure \ref{fig:theta_l_exp}) becomes nearly uniform at $t \approx 100$ s. The temperature drop in the gas is also significant and becomes nearly constant near the interface. At $t>150$s, after the mixing is complete, the temperature in the liquid increases again because of the exchanges with the wall. As this occurs, the pressure rises again, possibly due to evaporation as the interface warms up. This test case showcases the tremendous impact that sloshing-induced heat and mass transfer can have on the pressure and temperature of the ullage gas.

\begin{table*}[t]
	\centering
	\small
	\renewcommand{\arraystretch}{1.3}
	\caption{Optimal set of closure terms obtained after completing one hundred global optimizations in each test case. The uncertainties associated with each mean value are presented next to it, and were computed as two times the standard deviation.}
	\begin{tabular}{c|ccccccc}
		\hline
		Case                & $\overline{T}_w^0$ [K]                     & $h_{gi}^0$ [W/m2K] & $h_{li}^0$ [W/m2K] & $h_{wl}$ [W/m2K] & $h_{gi}^\infty$ [W/m2K] & $h_{li}^\infty$ [W/m2K] & $1/\tau$ [ms$^{-1}$] \\ \hline
		\multirow{3}{*}{s0} & $\overline{T}_l^0$                         &                    &                    & 25.82 ± 9.6      & 0.15 ± 0.01             & 0.1 ± 0                 &                      \\
		& $(\overline{T}_l^0   +\overline{T}_g^0)/2$ &                    &                    & 4.06 ± 0.05      & 0.14 ± 0.01             & 0.12 ± 0                &                      \\
		& $\overline{T}_g^0$                         &                    &                    & 2.01 ± 0.03      & 0.14 ± 0                & 0.12 ± 0                &                      \\ \hline
		\multirow{3}{*}{s1} & $\overline{T}_l^0$                         & 0.14 ± 0.03        & 0.25 ± 0.03        & 31.13 ± 11.5     & 0.11 ± 0.02             & 0.07 ± 0                & 13.4 ± 1.2           \\
		& $(\overline{T}_l^0   +\overline{T}_g^0)/2$ & 0.13 ± 0.02        & 0.26 ± 0.03        & 2.56 ± 0.21      & 0.12 ± 0.01             & 0.09 ± 0                & 14.2 ± 1             \\
		& $\overline{T}_g^0$                         & 0.13 ± 0.02        & 0.26 ± 0.04        & 1.26 ± 0.08      & 0.12 ± 0.01             & 0.09 ± 0                & 14.5 ± 1.4           \\ \hline
		\multirow{3}{*}{s2} & $\overline{T}_l^0$                         & 0.71 ± 0.2         & 2.97 ± 0.61        & 12.56 ± 7.93     & 0.23 ± 1.96             & 0.03 ± 0                & 34.6 ± 3.4           \\
		& $(\overline{T}_l^0   +\overline{T}_g^0)/2$ & 0.71 ± 0.15        & 2.98 ± 0.5         & 3.34 ± 0.87      & 0.12 ± 0.02             & 0.04 ± 0                & 34.1 ± 3.3           \\
		& $\overline{T}_g^0$                         & 0.7 ± 0.12         & 3.02 ± 0.45        & 1.65 ± 0.46      & 0.13 ± 0.01             & 0.04 ± 0                & 34.5 ± 3             \\ \hline
		\multirow{3}{*}{s3} & $\overline{T}_l^0$                         & 0.95 ± 0.14        & 2.88 ± 0.49        & 23.38 ± 10.27    & 0.14 ± 0.02             & 0.04 ± 0                & 35.9 ± 3.4           \\
		& $(\overline{T}_l^0   +\overline{T}_g^0)/2$ & 0.92 ± 0.13        & 2.84 ± 0.43        & 3.02 ± 0.66      & 0.14 ± 0.02             & 0.06 ± 0                & 35.5 ± 3.5           \\
		& $\overline{T}_g^0$                         & 0.86 ± 0.58        & 2.69 ± 0.9         & 1.59 ± 0.71      & 1.12 ± 5.92             & 0.06 ± 0                & 35.1 ± 4.2           \\ \hline
		\multirow{3}{*}{s4} & $\overline{T}_l^0$                         & 1.09 ± 0.13        & 3.95 ± 0.28        & 12.53 ± 7.44     & 0.1 ± 0.01              & 0.02 ± 0                & 41.1 ± 1.7           \\
		& $(\overline{T}_l^0   +\overline{T}_g^0)/2$ & 1.08 ± 0.14        & 3.95 ± 0.27        & 2.76 ± 0.79      & 0.1 ± 0.01              & 0.04 ± 0                & 40.8 ± 2             \\
		& $\overline{T}_g^0$                         & 1.05 ± 0.2         & 4.57 ± 3.63        & 1.46 ± 0.74      & 0.09 ± 0.05             & 0.1 ± 0.36              & 40.2 ± 5.2           \\ \hline
		\multirow{3}{*}{s5} & $\overline{T}_l^0$                         & 1.78 ± 1.58        & 6.36 ± 0.77        & 108.1 ± 37.14    & 3.84 ± 1.72             & 0 ± 0                   & 44 ± 5.8             \\
		& $(\overline{T}_l^0   +\overline{T}_g^0)/2$ & 2.64 ± 1.8         & 6.12 ± 0.69        & 25.41 ± 10.65    & 2.09 ± 1.91             & 0.1 ± 0.99              & 0.03 ± 0             \\
		& $\overline{T}_g^0$                         & 1.44 ± 1.03        & 5.92 ± 0.42        & 9.9 ± 0.62       & 3.1 ± 1.07              & 0 ± 0                   & 34.3 ± 2.7           \\ \hline
	\end{tabular}
	\label{tab:calibration_coeffs}
\end{table*}

\subsection{Data-driven discovery of heat transfer coefficients}
\label{sec:model_calibration}

% Introduction of the section. Tie it with the previous one, then introduce table 3 right away.
This section reports on the identification of the heat transfer coefficients through the inverse method and the accuracy of the resulting model's prediction. Table \ref{tab:calibration_coeffs} collects the main results for the terms in \eqref{eq:h_gi_exp} and \eqref{eq:h_li_exp} identified for the five test cases s0-s5 and the three scenarios for $\overline{T}_w(t=0)=\overline{T}_w^0$ described in Section \ref{sec:0d_model}. For each quantity, the table provides the mean value and the confidence interval, defined as twice the standard deviation, obtained from the columns of the $\mathbf{H}^*$ matrix in the Monte-Carlo evaluation (see Section \ref{sec:model_based_inverse_method}).

% As described in Section \ref{sec:0d_model}, the solid walls were treated as adiabatic boundaries in the model. Therefore, heat exchanges with these surfaces were not accounted for during the simulations. Although this may have contributed to underpredicting the rise in liquid temperature, especially in the long-term evolution of case s5, further research is still requi\subsection{}

\begin{figure*}[h]
	\centering
	\begin{subfigure}[b]{0.49\textwidth}
		\centering
		\includegraphics[width=\textwidth]{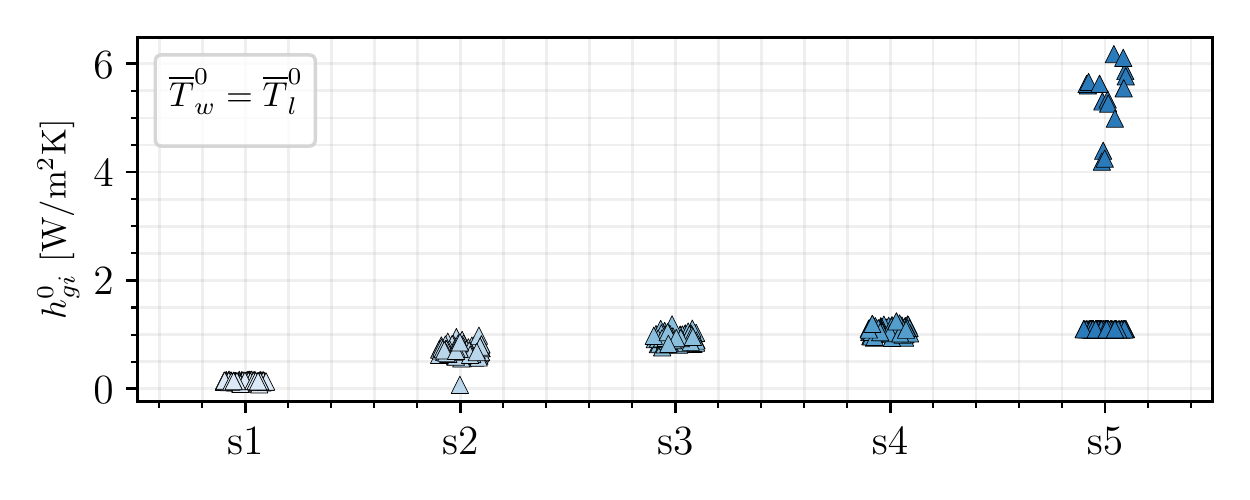}
		\caption{}    
		\label{fig:Liq_Tw_Hgi_swarm}
	\end{subfigure}
	\hfill
	\begin{subfigure}[b]{0.49\textwidth}  
		\centering 
		\includegraphics[width=\textwidth]{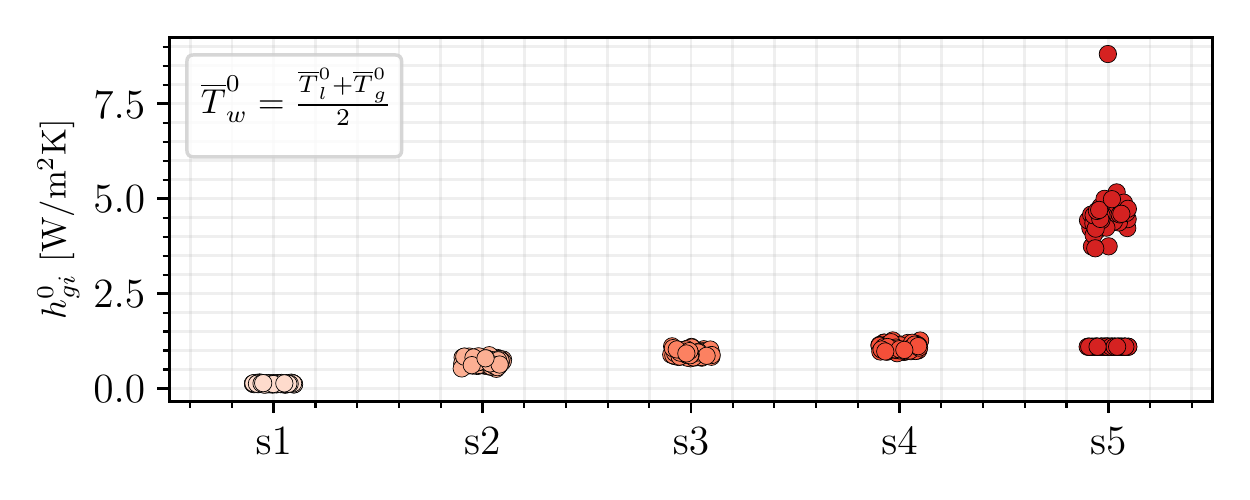}
		\caption{}    
		\label{fig:Avg_Tw_Hgi_swarm}
	\end{subfigure}
	\vskip\baselineskip
	\begin{subfigure}[b]{0.49\textwidth}   
		\centering 
		\includegraphics[width=\textwidth]{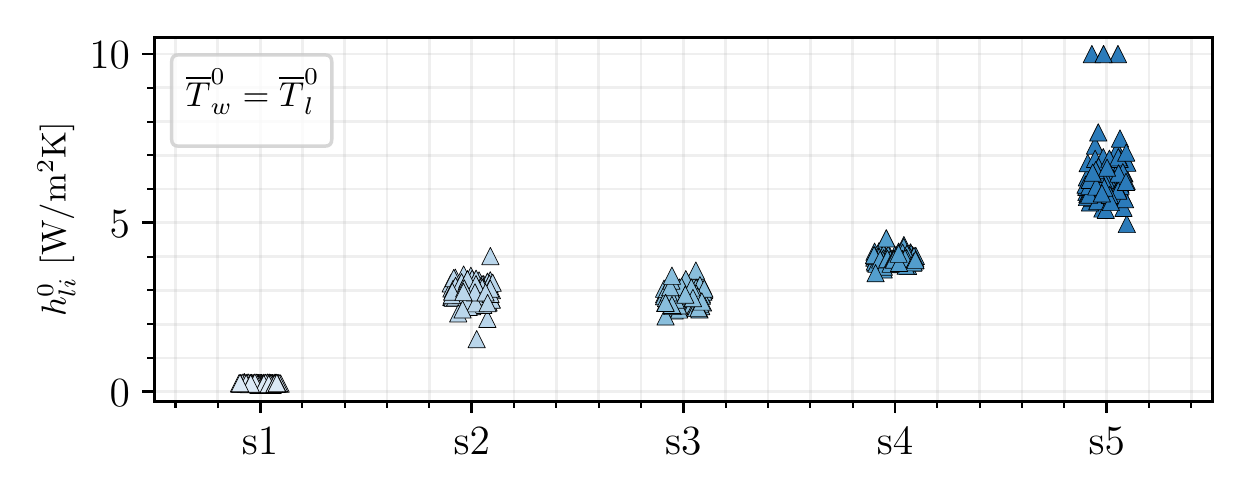}
		\caption{}    
		\label{fig:Liq_Tw_Hli_swarm}
	\end{subfigure}
	\hfill
	\begin{subfigure}[b]{0.49\textwidth}   
		\centering 
		\includegraphics[width=\textwidth]{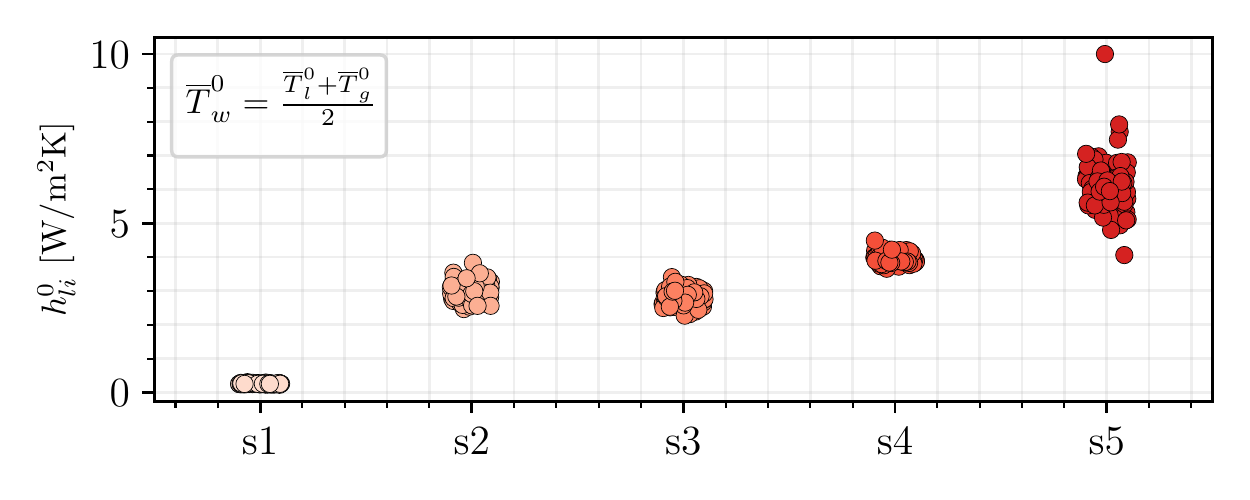}
		\caption{}   
		\label{fig:Avg_Tw_Hli_swarm}
	\end{subfigure}
	\caption{Swarm plots with the distribution of the optimal transient heat transfer coefficients (a) $h_{gi}^0$ with $\overline{T}_w^0=\overline{T}_l^0$, (b) $h_{gi}^0$ with $\overline{T}_w^0=(\overline{T}_l^0+\overline{T}_g^0)/2$, (c) $h_{li}^0$ with $\overline{T}_w^0=\overline{T}_l^0$, (b) $h_{li}^0$ with $\overline{T}_w^0=(\overline{T}_l^0+\overline{T}_g^0)/2$ for experimental test cases s1-s5.}
	\label{fig:H_swarmplot}
\end{figure*}

% Why only the transients with the interface? Explain...
Figure \ref{fig:H_swarmplot} complements this analysis by showing the distribution of the optimal $h_{gi}^0$, $h_{li}^0$ coefficients for test cases s1-s5. Case s0 was omitted because only $h_{gi}^\infty$, $h_{li}^\infty$ and $h_{wl}$ were calibrated in the absence of sloshing. 

%This was done because $h_{gi}$ and $h_{li}$ were expected to be constant in the absence of sloshing.
% And we had good matching with the data in these conditions

The results in Table \ref{tab:calibration_coeffs} demonstrate that the different scenarios for $\overline{T}_w^0$ have minimal impact on the coefficients discovered through the inverse method, except for $h_{wl}$. This parameter is maximum when $\overline{T}_w^0=\overline{T}_l^0$, and minimum when $\overline{T}_w^0=\overline{T}_g^0$. Interestingly, the inverse method adapts the optimal $h_{wl}$ to the wall temperature to provide the same heat exchange $\dot{Q}_{wl}(t)$. This could be expected from the fact that $h_{wl}$ and $\overline{T}_w(t)$ appear in the system of ODEs through the $\dot{Q}_{wl}(t)$ term. 
This is confirmed in Table \ref{tab:Q_wl}, which shows the time-averaged value of this heat flux for all experimental test cases and the three different wall temperature scenarios. The impact of the initial wall temperature on the predicted heat flux $\dot{Q}_{wl}$ is within 2.8\%, confirming its negligible influence on the regression of the other coefficients.

% Indeed, apart from $h_{wl}$, all other coefficients were within their uncertainty bounds regardless of the value chosen for $\overline{T}_w^0$. Therefore, 
%Although the true temperature of the walls was not measured experimentally, in reality, $\overline{T}_w^0$ must be between $\overline{T}_l^0$ and $\overline{T}_g^0$. 

%Thus, although we cannot provide insight into the expected values of $h_{wl}$ and $\overline{T}_w(t)$, we observe that most closure terms are nearly insensitive to this missing information.

%Although a perfect matching was not achieved, the average difference to the mean heat flux measured in each case was of 2.8\%, which was deemed enough to prove the insensitivity of the inverse method to the initial temperature of the wall.

\begin{table}[h]
	\centering
	\small
	\renewcommand{\arraystretch}{1.4}
	\caption{Average heat flux exchanged between the liquid volume and the tank walls $\dot{Q}_{wl}$ for test cases s0-s5, and for the three different wall temperature scenarios at $t=0$ s.}
	\begin{tabular}{c|ccc}
		\hline
		\multirow{2}{*}{Case} & \multicolumn{3}{c}{Time-averaged $\dot{Q}_{wl}$ [W]}                                                                                                \\
		& $\overline{T}_w^0=\overline{T}_l^0$ & $\overline{T}_w^0=\frac{\overline{T}_l^0   +\overline{T}_g^0}{2}$ & $\overline{T}_w^0=\overline{T}_g^0$ \\ \hline
		s0                    & 0.72                                & 0.72                                                              & 0.74                                \\
		s1                    & 0.42                                & 0.42                                                              & 0.43                                \\
		s2                    & 0.58                                & 0.50                                                              & 0.54                                \\
		s3                    & 0.49                                & 0.50                                                              & 0.53                                \\
		s4                    & 0.43                                & 0.45                                                              & 0.50                                \\
		s5                    & 2.84                                & 2.82                                                              & 2.89                                \\ \hline
	\end{tabular}
	\label{tab:Q_wl}
\end{table}

% Comment on the evolution of the transient heat transfer coeffs
% We now proceed to analysing the other heat transfer coefficients identified through the inverse method.
From Table \ref{tab:calibration_coeffs} and Figure \ref{fig:H_swarmplot}, it is evident that $h_{li}^0$, $h_{gi}^0$, and $h_{wl}$ increase with the sloshing-based Reynolds number. This is consistent with the pressure and temperature evolutions reported in the previous section. As expected, the 0D model correlates the sudden pressure and the temperature drops observed in Figures \ref{fig:pg_all_exp}-\ref{fig:theta} for cases s2-s5 with larger values of $h_{li}^0$ and $h_{gi}^0$ than those identified in cases s1 and s0. On the other hand, he identified $h_{gi}^\infty$ is nearly the same in all cases s0-s4. This suggests that the heat exchanges between the gas and the interface, at large times, are independent of the sloshing conditions. These results agree with the experimental measurements presented in Section \ref{sec:p_T_exp}, which show that planar sloshing triggers the initial thermal destratification, but the long-term evolution is driven by conduction because the convective fluxes cannot reach the bulk fluid.

This result also aligns with the identified decay time $\tau$, reported in the last column, Table \ref{tab:calibration_coeffs}. As highlighted in Section \ref{sec:5p2}, the transient sloshing dynamic is characterized by a decay rate $\omega_{1,1}\gamma_{1,1}=48.7$ ms$^{-1}$, which is comparable to the time scale $\tau$ for the cases s3-s5, where the pressure drop was particularly evident.
This suggests that the time scale of the de-stratification is comparable to the time scale of the sloshing's transient dynamics. Moreover, it is possible that the derived coefficients are not only linked to the sloshing regime per se (planar or swirl), but also the transient sloshing dynamics.

%Thus, it is possible that under the presented formulation, the model's closure terms might depend not only on the dominant heat transfer mechanism (i.e., sloshing-induced forced convection, free convection, or conduction), but also on the transient motion of the interface. Indeed, since the heat transfer coefficients depend on the convective fluxes present in the flow, the degree of thermal mixing promoted during the transient sloshing period must be different than in the harmonic regime.

%In line with the results presented in Figures \ref{fig:pg_all_exp}-\ref{fig:theta}, while cases s0 and s1 are characterized by gradual pressure drops and negligible thermal destratification, cases s2-s5 are characterized by steep pressure variations, coupled with significant mixing at the start of sloshing. The sudden changes in pressure and temperatures are well correlated, by the 0D model, to larger values of $h_{li}^0$ and $h_{gi}^0$.

% is responsible for the initial thermal destratification until the temperature distribution becomes more homogeneous. After this point, because the convective fluxes cannot reach the bulk flow, the thermal information must be transported via conduction.
% Although the steady-state responses of s0-s4 were similar, the transients were distinct due to different heat transfer coefficients and $\tau$ constants between these test cases. In particular, $\tau$ was found to decrease as $\text{Re}_s$ increased, meaning shorter transient periods were associated with stronger thermal mixing.

Figure \ref{fig:H_swarmplot} shows that the distribution of identified coefficients is narrow and unimodal for all cases except for case s5. This test case produced the steepest pressure drop, leading to homogeneous temperature fields in the liquid after 30 excitation periods. Accordingly, the inverse method derives the highest $h_{gi}^0$ and $h_{li}^0$ coefficients. Moreover, it predicts the largest amount of heat exchanged with the walls (see Table \ref{tab:Q_wl}), which appears much larger than the one exchanged with the interface ( $\dot{Q}_{li}$ ). Therefore, the model suggests that the rise in the liquid temperature in Figure \ref{fig:theta_l_exp} is primarily due to the heat transfer with the solid walls.

\begin{figure*}[h]
	\centering
	\begin{subfigure}[b]{0.49\textwidth}
		\centering
		\includegraphics[width=\textwidth]{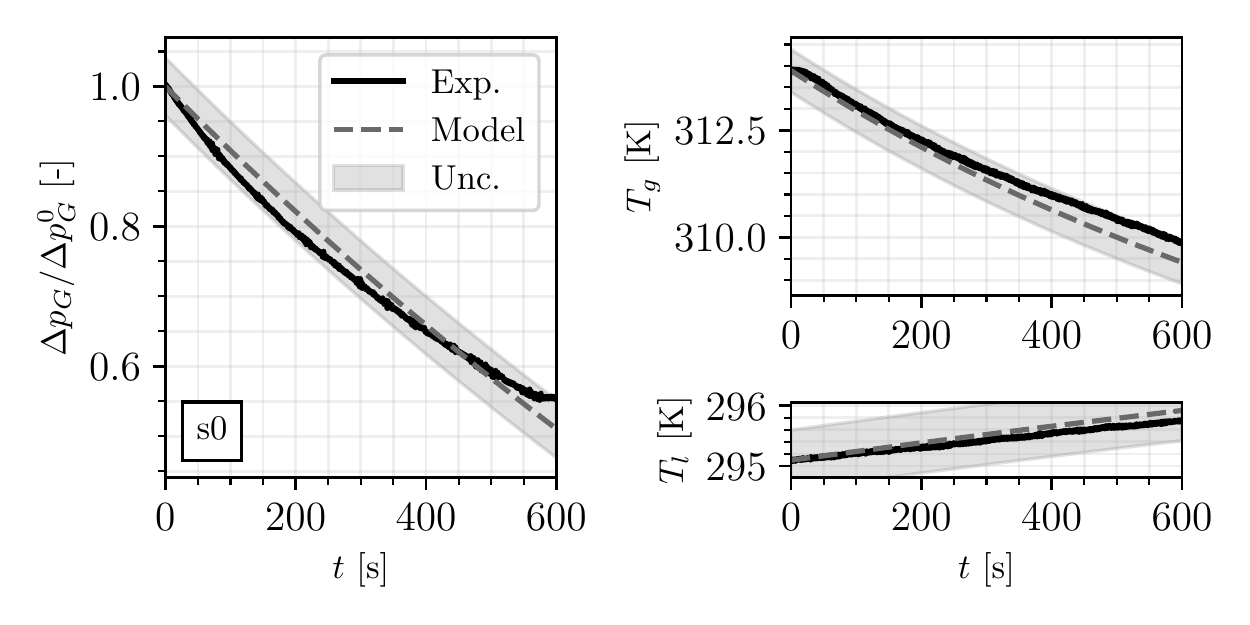}
		\caption{Case s0: system relaxation in quiescent conditions}    
		\label{fig:s0_results}
	\end{subfigure}
	\hfill
	\begin{subfigure}[b]{0.49\textwidth}  
		\centering 
		\includegraphics[width=\textwidth]{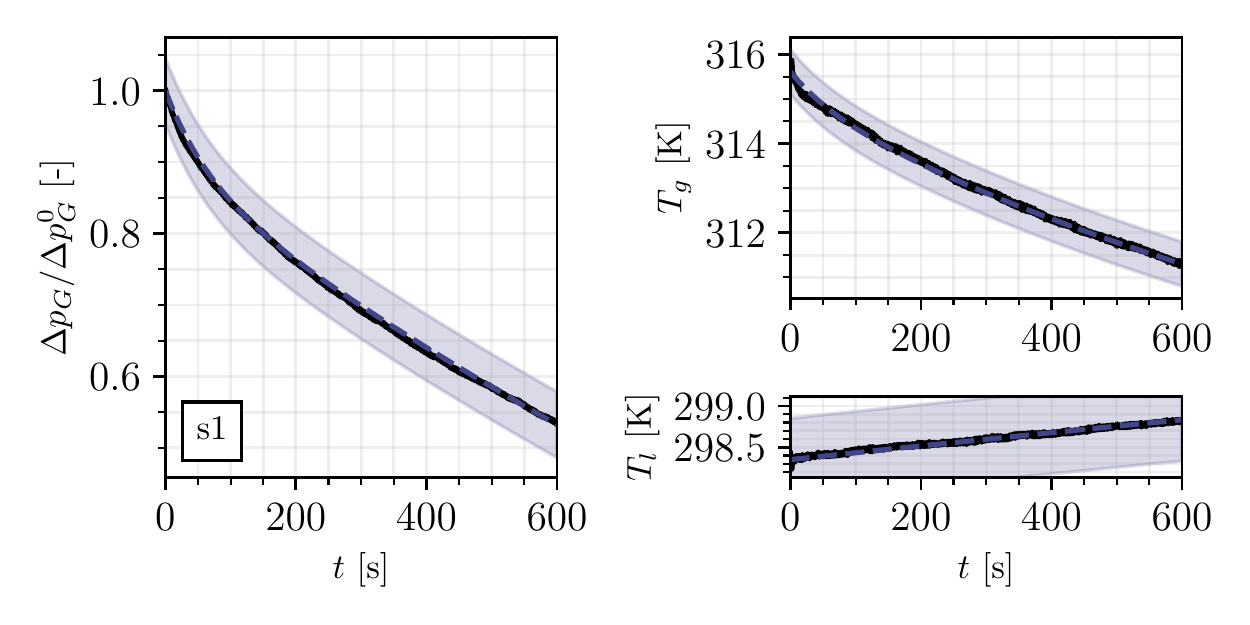}
		\caption{Case s1: subcritical planar sloshing}    
		\label{fig:s1_results}
	\end{subfigure}
	\vskip\baselineskip
	\begin{subfigure}[b]{0.49\textwidth}   
		\centering 
		\includegraphics[width=\textwidth]{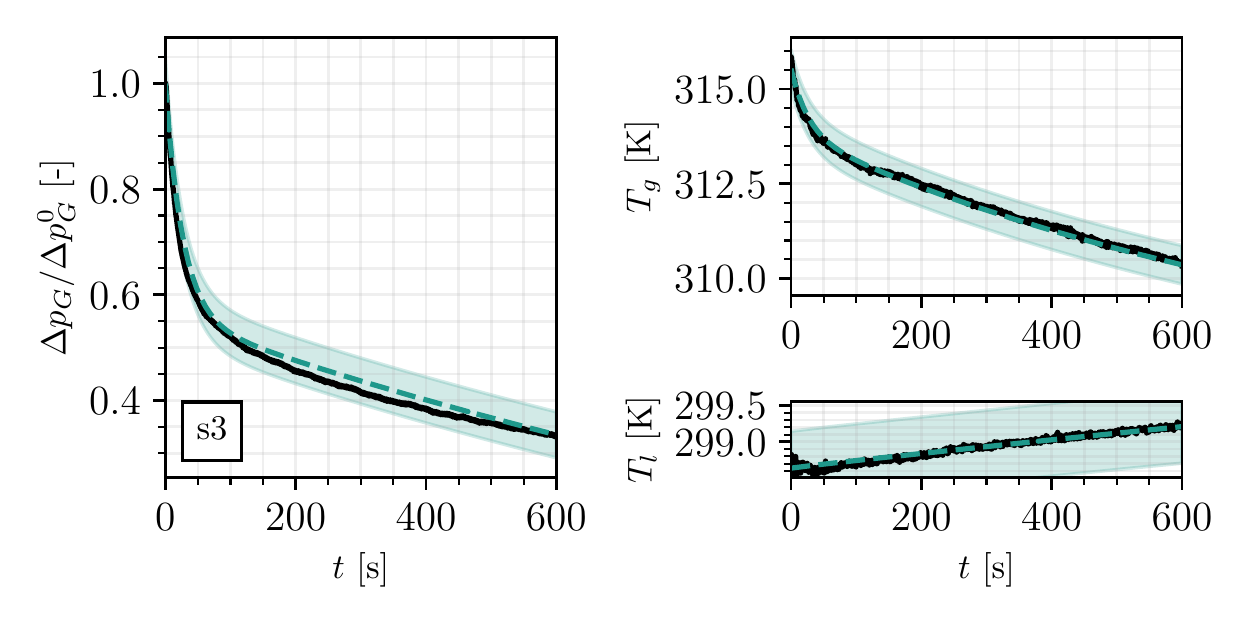}
		\caption{Case s3: supercritical planar sloshing}    
		\label{fig:s3_results}
	\end{subfigure}
	\hfill
	\begin{subfigure}[b]{0.49\textwidth}   
		\centering 
		\includegraphics[width=\textwidth]{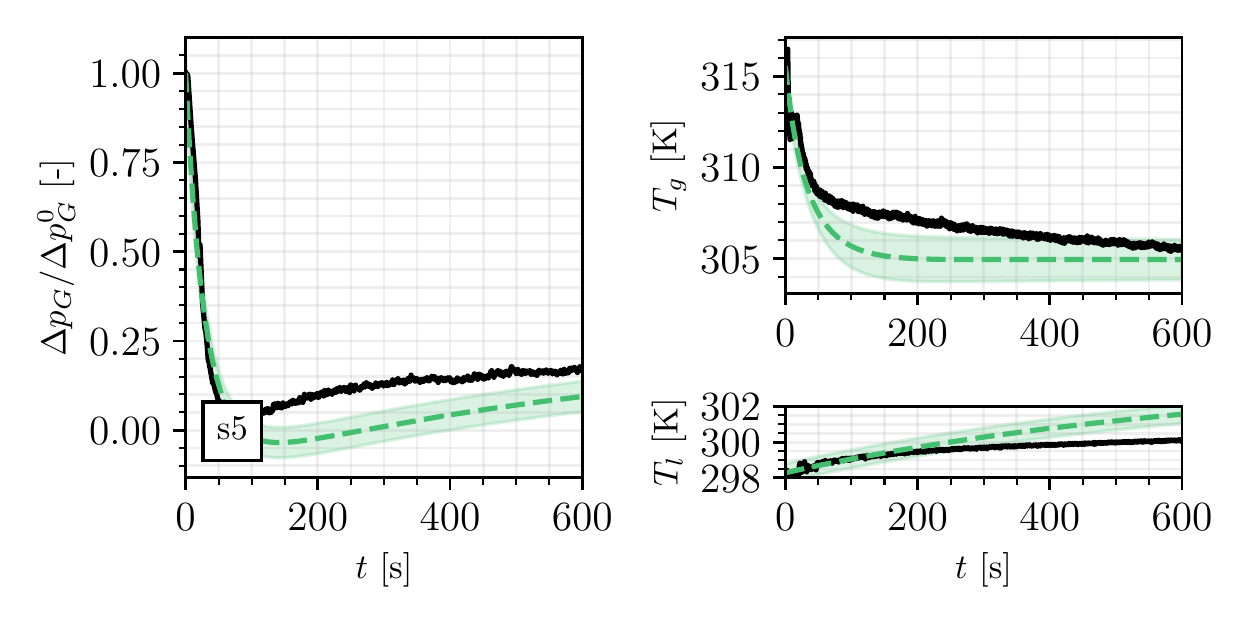}
		\caption{Case s5: swirl sloshing}   
		\label{fig:s5_results}
	\end{subfigure}
	\caption{Results of the 0D model calibration applied to test cases s0, s1, s3 and s5, considering the initial wall temperature $\overline{T}_w^0=(\overline{T}_l^0+\overline{T}_g^0)/2$. The data was generated using the mean closure terms from Table \ref{tab:calibration_coeffs}, and the uncertainty region was computed via a Monte Carlo approach (see Section \ref{sec:model_based_inverse_method}).}
	\label{fig:0D_results}
\end{figure*}

The results of the 0D model are compared against the experimental measurements for cases s0, s1, s3 and s5 in Figure \ref{fig:0D_results}, which was generated for the scenario $\overline{T}_w^0=(\overline{T}_l^0+\overline{T}_g^0)/2$. The dashed lines represent the pressure and temperature evolutions generated using the mean $\mathbf{\bar{h}^*}$ coefficients from Table \ref{tab:calibration_coeffs}, and with the initial conditions of each test case $\mathbf{X}^0$. The shaded area represents the uncertainty of the prediction, estimated as described in Section \ref{sec:model_based_inverse_method}.

The results of the calibrated model showed good agreement with the experimental data, which was generally within the uncertainty bands of the prediction. Figures \ref{fig:s0_results} and \ref{fig:s1_results} correspond to tests s0 and s1, respectively. While the pressure and gas temperature evolutions were nearly identical in these tests, steeper initial drops were registered for s1. This observation aligns with the much lower value of $h_{gi}^0$ in case s0 compared to s1. This difference is likely due to to the sudden excitation applied to case s1, which reduced the response time of the system compared to s0.
Since the optimal coefficients in s0 were determined without the transient terms on the interface, the temperature drop in the ullage is overpredicted by 11\%, whereas the numerical pressure drop is 5\% more than the experimental one.
On the other hand, the matching of case s1 is remarkable, with the mean numerical evolutions overlapping the experimental ones. 
The supercritical planar sloshing tests are represented by case s3 in Figure \ref{fig:s3_results}. This figure highlights the model's ability to capture the system's transient pressure and temperature variations. Similar to s1, there is nearly perfect agreement between the numerical and experimental curves, which is remarkable given the steep initial variation in pressure and temperature.

Finally, Figure \ref{fig:s5_results} concerns the swirl test case s5. While the transient evolution of the system was well captured by the model, the following pressure rise was not well reproduced. Moreover, the rise in liquid temperature was overestimated by nearly two degrees, while the decrease in ullage temperature was steeper in the simulations. Nevertheless, the model showed promising agreement with the experiments, especially given its simple formulation and negligible computational cost.

\begin{figure*}[htb!]
	\centering
	\begin{subfigure}[b]{0.49\textwidth}
		\centering
		\includegraphics[width=\textwidth]{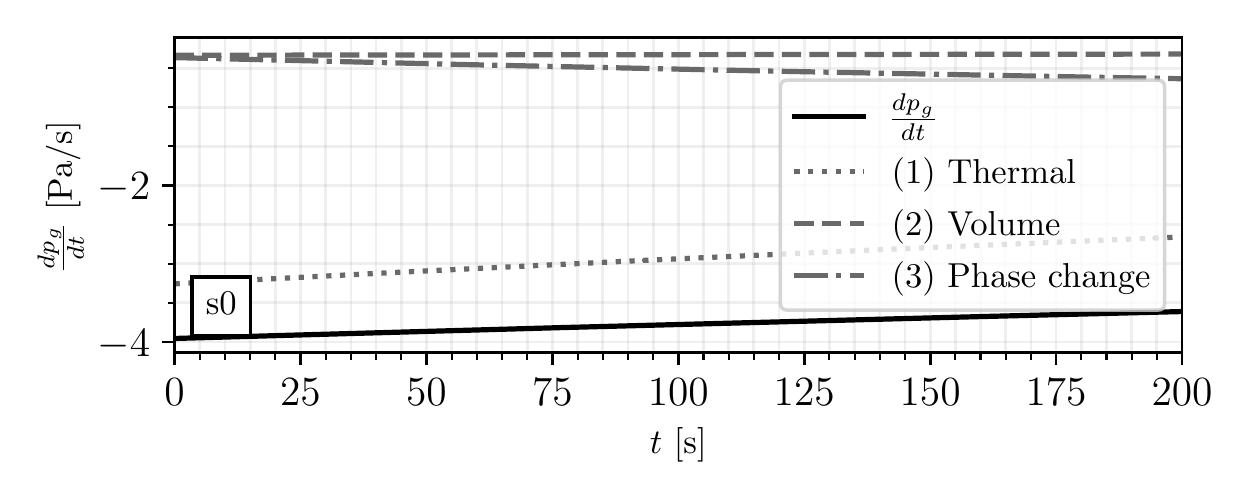}
		\caption{}    
		\label{fig:dpgdt_s0}
	\end{subfigure}
	\hfill
	\begin{subfigure}[b]{0.49\textwidth}  
		\centering 
		\includegraphics[width=\textwidth]{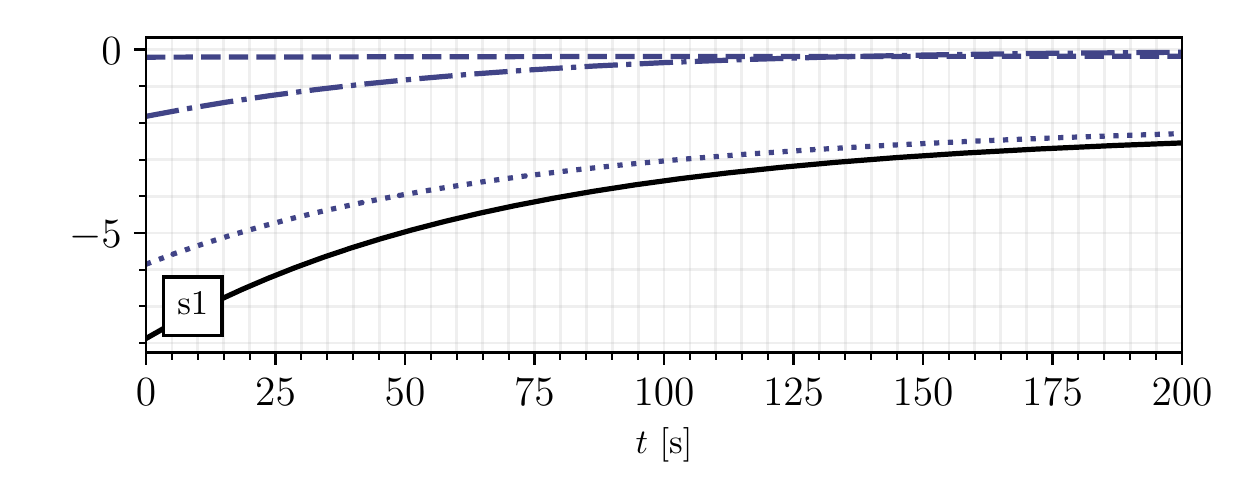}
		\caption{}    
		\label{fig:dpgdt_s1}
	\end{subfigure}
	\vskip\baselineskip
	\begin{subfigure}[b]{0.49\textwidth}   
		\centering 
		\includegraphics[width=\textwidth]{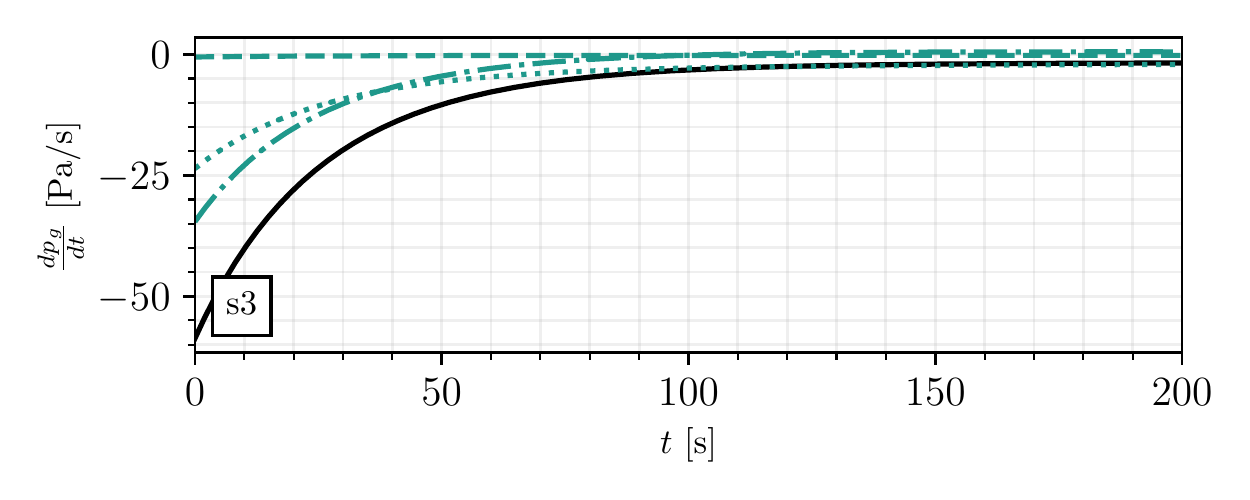}
		\caption{}    
		\label{fig:dpgdt_s3}
	\end{subfigure}
	\hfill
	\begin{subfigure}[b]{0.49\textwidth}   
		\centering 
		\includegraphics[width=\textwidth]{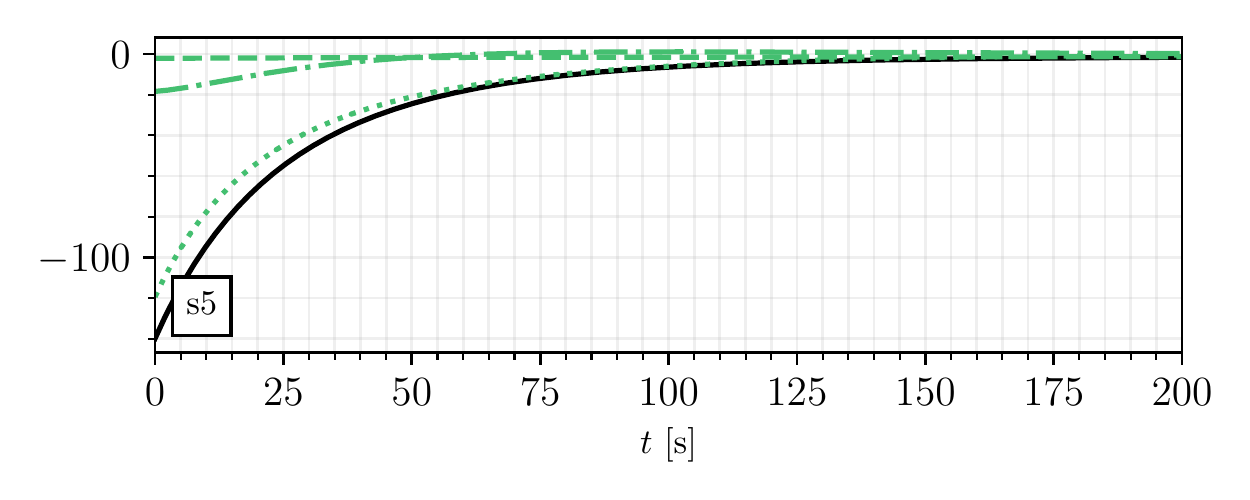}
		\caption{}   
		\label{fig:dpgdt_s5}
	\end{subfigure}
	\caption{Pressure drop rate due to variations of the ullage (1) temperature, (2) volume, and (3) mass identified in Equation \eqref{eq:pressure_evolution} for cases (a) s0, (b) s1, (c) s3, and (d) s5. These individual contributions are also compared to the overall rate-of-change of the ullage pressure, drawn in solid black lines.}
	\label{fig:dpgdt_all}
\end{figure*}

Lastly, the model was used to assess the relative importance of the different terms which drive the pressure evolution according to Equation \eqref{eq:pressure_evolution}. 
Figure \ref{fig:dpgdt_all} shows the different contributing terms in $dp_g/dt$, computed with $\overline{T}_w^0=(\overline{T}_l^0+\overline{T}_g^0)/2$, for $t\in[0,200]$ s.
Solid lines represent the overall pressure change-rate, dotted lines correspond to the variation due to thermal expansion/contraction, dashed lines represent the term accounting for changes in volume, and dash-dotted lines indicate the contribution of adding or removing vapour mass due to phase change through the interface.
This last term depends on $dm_v/dt$, which depends on the mass transfer coefficient defined in Equation \eqref{eq:mass}. %Although this parameter was not directly used as a closure term for the model, its magnitude is indirectly evaluated in this section through its contribution to the pressure variations.
These figures show that the volumetric term can be neglected in all test cases. 
%This was expected since only minimal changes in the fill level were observed in the experimental campaign.

On the other hand, the relative weight of the other two terms changes significantly with the sloshing conditions. The pressure drop in cases s0 and s1 is dominated by the thermal (cooling) contribution. Mass transfer and cooling have comparable weight in case s3, while case s5 is largely dominated by the mass transfer (condensation) effect until $t\approx 40$s.
Approximately 85\% of the initial pressure drop in s5 is due to condensation. Afterwards, the pressure evolution is again driven by the thermal effect, and a change of sign is observed at $t > 60$ s. In this last phase, the liquid's temperature homogenized (see Figure \ref{fig:theta_l_exp}) and started warming up because of heat exchanges with the walls. This triggers evaporation, hence a rise in the tank's pressure.

\section{Conclusions}
\label{sec:conclusion}

% Summary of the goals and what was done

We presented an experimental characterization of non-iso\-thermal sloshing in an upright cylindrical tank under lateral harmonic excitation and proposed a model-based inverse method to infer heat and mass transfer coefficients.

The experiments were carried out on the Karman Institute's SHAKESPEARE sloshing table in a reduced-scale quartz tank filled with HFE-7200 and instrumented with an array of thermocouples and a pressure tap. In all sloshing experiments, the tank is first pressurized and thermally stratified using heating elements on the ullage gas side. Temperature distribution and pressure were monitored both in the preparatory phase and the sloshing experiments, and Particle Image Velocimetry (PIV) was used to prove that buoyant flows have a negligible impact on the liquid. Hence, the initial conditions for the sloshing are those of a quiescent and thermally stratified. Four main scenarios were analyzed: (1) absence of sloshing, (2) subcritical planar waves (2) subcritical planar waves, (3) supercritical planar waves, and (4) swirl waves. 

The temperature and pressure measurements were used to infer heat and mass transfer coefficients in the tank using a model-based inverse method. These are identified via an optimization, carried out with the Basinhopping algorithm, as the one that minimizes the discrepancy between data and model prediction. The model is a simple 0D formulation of heat and mass balance between liquid, gas and walls.

In line with previous experimental works, it was observed that the subcritical planar waves (Re$_s$ $<$ (Re$_{s}$)$_c$) do not appreciably increase the heat and mass transfer from the quiescent conditions. Instead, supercritical planar waves promote thermal mixing in the vicinity of the gas-liquid interface and produce an appreciable pressure drop in the tank. The mixing and the associated pressure drop are considerably larger in swirling conditions, which destroy the thermal stratification.

In all cases of sloshing enhanced heat and mass transfer rate, the model predicts that the pressure drop is primarily due to condensation from the ullage gas due to a change in temperature of the gas-liquid interface.

Considering the important simplifications underpinning the 0D model used in the inverse method, the results are particularly promising. If correctly calibrated, the model could reproduce the experimental data accurately in all the investigated conditions at a negligible computational cost. These results thus open the path to using model-based inverse derivations of closure laws to describe the thermodynamic evolution of non-isothermal sloshing. The same approach could be extended to all critical phases of cryogenic fuel storage (tank filling, pressurization, and long-term storage). Besides allowing for finding correlations from real-time data (for example, by searching for a link between the closure coefficients and states of the system), these could also be used for model predictive control and anomaly detection in thermal management systems. Both applications are currently under investigation.

\section*{Acknowledgments}
This  work was supported  by the European  Space Agency (ESA) in  the  framework of the GSTP-SLOSHII project with reference number 4000129315/19/NL/MG. The authors gratefully acknowledge the financial support of the `Fonds de la Recherche Scientifique (F.R.S. -FNRS)' for the FRIA grant with reference 40009348 supporting the PhD of Mr. Marques.

\bibliographystyle{elsarticle-num} 
\bibliography{Marques_et_al_2022}

\end{document}